\documentclass[aps,jap,preprint,amsmath,floatfix,amssymb]{revtex4}

\newcommand{\be}{\begin{equation}}
\newcommand{\ee}{\end{equation}}
\newcommand{\ba}{\begin{array}}
\newcommand{\ea}{\end{array}}
\newcommand{\bea}{\begin{eqnarray}}
\newcommand{\eea}{\end{eqnarray}}
\newcommand{\beann}{\begin{eqnarray*}}
\newcommand{\eeann}{\end{eqnarray*}}
\newcommand{\bn}[1]{\mbox{\boldmath $#1$}}

\def\mb{\mbox}

\def\p{^{\;\prime}}

\usepackage{latexsym}
\usepackage[dvips]{graphicx}
\usepackage{dcolumn}
\usepackage{amsmath}
\usepackage[latin1]{inputenc}
\usepackage{graphicx}
\usepackage{color}
\usepackage{amsxtra}

\begin{document}

\title{Quantum-ring spin interference device tuned by  quantum point contacts}

\author{Leo Diago-Cisneros$^{1}$}
\author{Francisco  Mireles$^{2}$}
\affiliation{$^{1}$Facultad de F\'{\i}sica, Universidad de La Habana, C.P.10400, La Habana, Cuba }
\affiliation{$^{2}$Centro de Nanociencias y Nanotecnolog\'{\i}a, Universidad Nacional Aut\'{o}noma de M\'{e}xico, C.P. 22800 Ensenada, Baja California, M\'exico.}

\begin{abstract}
We introduce a spin-interference device that comprises a quantum ring (QR) with three embedded quantum point contacts (QPCs) and study theoretically its spin transport properties in the presence of Rashba spin-orbit interaction. Two of the QPCs conform the lead-to-ring junctions while a third one is placed symmetrically in the upper arm of the QR. Using an appropriate scattering model for the QPCs and the $\mathbb{S}$-matrix scattering approach, we analyze the role of the QPCs on the Aharonov-Bohm (AB) and Aharonov-Casher (AC) conductance oscillations of the QR-device. Exact  formulas are obtained for the spin-resolved conductances of the QR-device as a function of the confinement of the QPCs and the AB/AC phases. Conditions for the appearance of resonances and anti-resonances in the spin-conductance are derived and discussed. We predict very distinctive variations of the QR-conductance oscillations not seen in previous QR proposals.  In particular we find that the interference pattern in the QR can be manipulated to a large extend by varying electrically the lead-to-ring topological parameters. The latter can be used to modulate the AB  and AC phases by applying gate voltage only. We have shown also  that the conductance oscillations exhibits a crossover to well-defined resonances as the lateral QPC confinement strength is increased, mapping the eigenenergies of the QR. In addition,  unique features of the conductance arises by varying the aperture of the upper-arm QPC  and the Rashba spin-orbit coupling.  Our results may be of relevance for promising spin-orbitonics devices based in quantum interference mechanisms.

\end{abstract}


\date{\today}
\maketitle

\section{Introduction}
 \label{Sec:Intro}

Quantum rings structures not only provides an excellent physics playground to study coherent transport and electronic interference phenomena, but they also constitutes very appealing systems towards the realization of novel spintronic\textcolor[rgb]{0.00,0.00,1.00}{\cite{Zutic}} devices. The Aharonov-Bohm\textcolor[rgb]{0.00,0.00,1.00}{\cite{AB59}} (AB) effect and its relativistic counterpart, the Aharonov-Casher\textcolor[rgb]{0.00,0.00,1.00}{\cite{AC84}}(AC) effect, are just two of these quantum interference phenomena that may appear simultaneously in semiconductor quantum rings. They manifest as an oscillatory behavior of the conductance of the quantum ring, either as a magnetic flux is varied (AB-effect), or as a function of the spin-orbit interaction (SOI) strength (AC-effect) in the semiconductor ring. In the AB-effect, is the accumulated phase acquired by two electronic waves traveling coherently around a close loop in opposite directions that leads to magnetoconductance oscillations.  In contrast, in the AC-effect, is the precessing intrinsic magnetic moment (spin) of the electrons winding the ring, that acquires a phase change due to a gate-voltage tunable SOI.\textcolor[rgb]{0.00,0.00,1.00}{\cite{Nitta97,Engels,Schapers,Vasilopoulos05}} As a result, periodic conductance oscillations can be observed as a   perpendicular external electric field is varied.\textcolor[rgb]{0.00,0.00,1.00}{\cite{Nitta09}}

The AB-effect was first observed by Webb {\it et al.} in metallic rings.\textcolor[rgb]{0.00,0.00,1.00}{\cite{Webb}}A few years later Cimmino {\it et al.}\textcolor[rgb]{0.00,0.00,1.00}{\cite{Cimmino}} performed experiments using neutron (spin 1/2) beams following a closed path observing signals of the AC-oscillation phenomena. The AC-phase oscillations has been observed also in a HgTe/HgCdTe based single quantum ring\textcolor[rgb]{0.00,0.00,1.00}{\cite{Konig}} and in small arrays of mesoscopic  InGaAs/InAlAs based rings\textcolor[rgb]{0.00,0.00,1.00}{\cite{Bergsten}} which exhibit strong Rashba-SOI.\textcolor[rgb]{0.00,0.00,1.00}{\cite{Rashba}}

Additional phase effects concern the Berry (geometric) phase, acquired by a quantal particle traveling around a circuit in inhomogeneous effective magnetic fields.\textcolor[rgb]{0.00,0.00,1.00}{\cite{Berry84}} Berry phase signatures in the conductance has been studied in one- and two-dimensional (2D) rings with Rashba SOI.\textcolor[rgb]{0.00,0.00,1.00}{\cite{Frustaglia04,Frustaglia05}} More recently the AC-phase oscillations were measured in an array of gated InGaAs/InAlAs-based quantum rings as a function of the Rashba SOI strength and of the rings radius.\textcolor[rgb]{0.00,0.00,1.00}{\cite{Nitta12}} Interestingly, the authors were also able to observe a topological (geometric) spin phase,\textcolor[rgb]{0.00,0.00,1.00}{\cite{Qian}} that together with its dynamical phase, contributes to the overall time-reversal AC-phase change.\textcolor[rgb]{0.00,0.00,1.00}{\cite{Nitta12,KlausR}}

Micron-sized and nanoscale semiconductor quantum rings fabricated on two-dimensional electron gases (2DEG) are the base of several spintronics proposal devices, ranging from spin-interference devices that modulate the output electric current,\textcolor[rgb]{0.00,0.00,1.00}{\cite{Nitta99,Aronov}} spin-filters, \textcolor[rgb]{0.00,0.00,1.00}{\cite{Kiselev,ShelykhPRB72}} and quantum splitters,\textcolor[rgb]{0.00,0.00,1.00}{\cite{ShelykhPRB71}} to mention a few.

In this work we propose a spin-interference device setup composed by a semiconductor quantum-ring (QR) with Rashba-SOI at the center of the structure, and connected to a source and drain of electrons via quantum point contacts (QPCs). In contrast to previous studies, the ring junctions (here characterized by QPCs) are allocated symmetrically, just at the QR periphery. Additionally a third QPC is located in one of the arms of the QR device (Fig.\ref{QR-device}). Most notably, all QPCs are presumed to be 2D and independently tuned with in-plane electrical gates. By doing so, we additionally manage to overcome several disadvantages of former related studies that relied on a free parameter $\epsilon$ to characterize the QR-coupling with the leads.\textcolor[rgb]{0.00,0.00,1.00}{\cite{Aronov,Shelykh05,Vasilopoulos07}}

As far we know, a study of the role of QPCs with variable electrostatic width,  allocated at the inlet and outlet of a quantum ring (QR), has not been reported before, neither experimentally, nor theoretically. Same applies for the case of a QR-device with an inserted QPC with tunable width at one of the arms of the QR.  Hence just the proposal itself constitutes an interesting idea worth of a detailed study. Is very important to remark also that this setup offers the possibility of tuning the spin-interference phenomenon in the QR just by controlling electrostatically the effective width of the QPCs, in the presence or not, of an external magnetic fields as we shall see later.

Here we study theoretically the coherent electronic quantum transport and the spin interference in the described QR-based setup above in the presence of Rashba-SOI, and investigate the role of the lead-to-ring junctions via QPCs together with an upper-arm QPC. The backscattering and tunneling --through the lead-to-ring junctions-- were modeled by two-dimensional saddle point potentials describing here the QPC electrical confinement. Following B\"uttiker\textcolor[rgb]{0.00,0.00,1.00}{\cite{Buttiker84}}and Vasilopoulus {\it et al.}\textcolor[rgb]{0.00,0.00,1.00} {\cite{Vasilopoulos07}} an appropriate scattering matrix $\mathbb{S}$ formalism is established. Invoking then time-reversal invariance and unitarily relations of the $\mathbb{S}$ matrix we are able to derive the relevant tunneling quantities through the whole structure. Exact formulas for the quantum conductance of the QR-device are obtained. We find that two-probe conductance oscillations of the ring are rather sensitive to the confinement strength at the QPCs, as well as with the applied magnetic field and the spin orbit interaction intensity. We predict distinctive variations of the two-probe QR-conductance oscillations which can be used to modulate the Aharonov-Bohm and Aharonov-Casher phases by applying gate voltages. It is shown as well that is possible to control interference pattern in the QR to a large extend by varying the lead-to-ring topological parameters. Our results might be therefore of relevance for promising spin-orbitonics\textcolor[rgb]{0.00,0.00,1.00}{\cite{KlausR}}  devices based in quantum interference mechanisms.

The remainder of the  paper is organized as follows. In Section \ref{Sec:Device} we introduce the quantum ring device under study as well as the QPC model employed. In Section \ref{QR-Hamiltonian} we analyze the bare quantum ring Hamiltonian and its eigensolutions.  The $\mathbb{S}$-matrix formalism applied to the whole QR-device is presented in Section \ref{Sec:Scattering}. The main formulas of this paper are derived in this section. In Section \ref{Sec:Results} we discuss in detail  several illustrative numerical simulations of the spin interference properties occurring in the proposed device configuration. A Summary is provided in Section \ref{Sec:Summary}.

\section{Quantum Ring Device Proposal and the \textcolor[rgb]{0.00,0.00,1.00}{2D}-QPC model }
 \label{Sec:Device}
\subsection{Quantum ring device geometry}

\begin{figure}
\hspace*{-0.1cm}{\includegraphics[width=3.6in,height=2.2in]{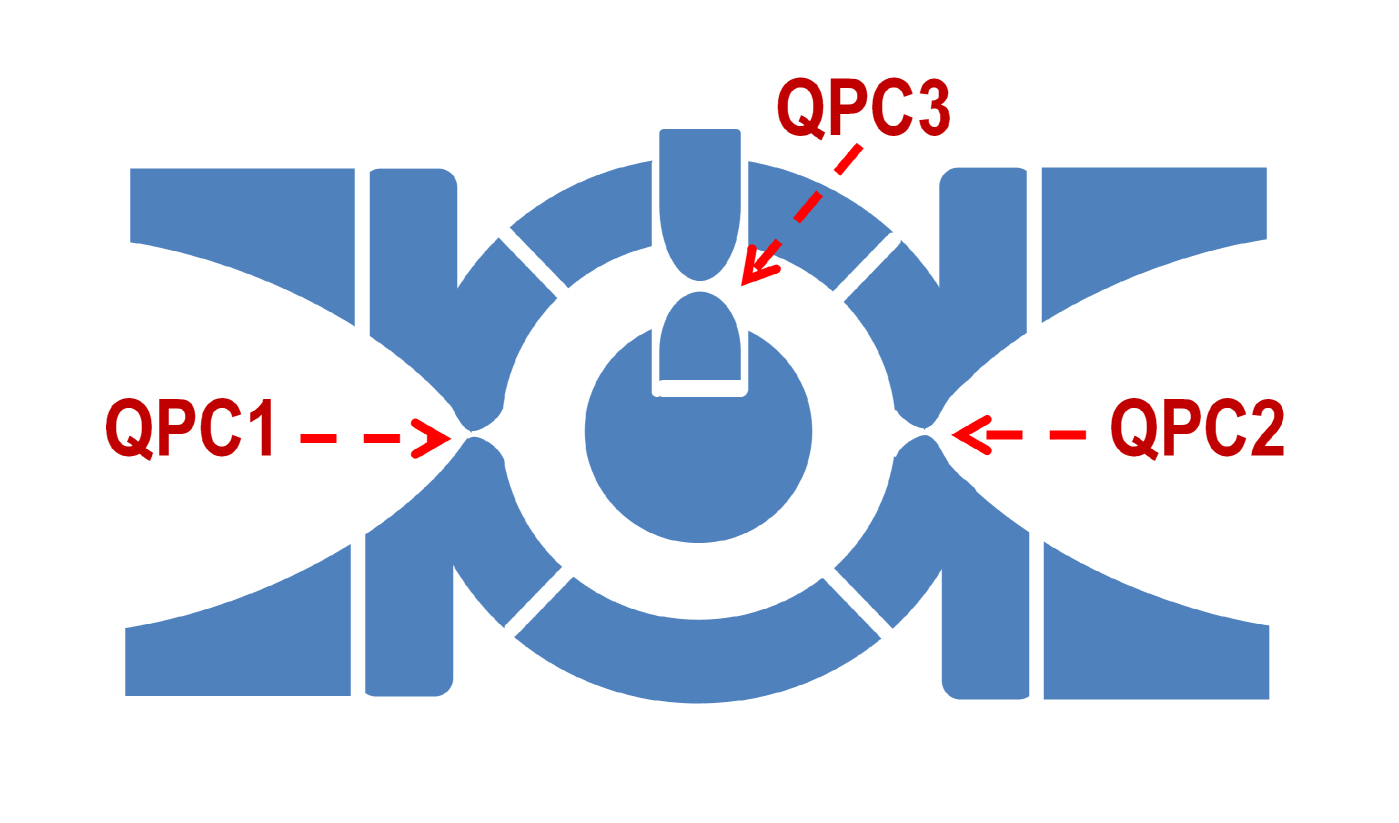}}
\vspace{-0.3in}
\caption{\label{QR-device} (Color online) Schematic diagram showing a top view of the envisioned QR device. Two quantum point contacts (QPC$_{1,2}$) are allocated symmetrically outside the QR path, just at the QR periphery. An additional QPC$_{3}$ is considered in the QR upper arm. All QPCs are supposed to be controlled independently by in-plane gate voltages. }
\end{figure}

Consider a quantum ring of radius $R$ and width $w$ defined electrostatically on a semiconductor 2DEG through  some radial confining potential. It can be produced {\it e.g.} by applying a gate voltage to a metallic ring-shape  structure deposited on top of the semiconductor heterostructure  containing the 2DEG. Fig.\,\,\ref{QR-device} shows a sketch of a top view of the whole device configuration envisioned here.  We assume that the two-probe device consist of the QR at the center of the structure connected to two leads by QPCs conforming the lead-to-ring junctions (interfaces). The leads are considered in ohmic contact with a QPC/QR$^{*}$/QPC structure, where QR$^{*}$ denotes here that an additional QPC has been inserted in the upper arm of the QR. The effective widths (transparency) of all QPCs are assumed to be gate-tunable in an independent manner.

Theoretically, the role of the electron scattering at the junctions lead-to-ring on the Aharonov-Bohm oscillations was first addressed by B\"uttiker\textcolor[rgb]{0.00,0.00,1.00}{\cite{Buttiker84}} by studying metallic QRs within the $\mathbb{S}$-matrix formalism. The approach has been employed subsequently to model transport in semiconductor AB-rings as well as AC spin-orbit effects.\textcolor[rgb]{0.00,0.00,1.00}{\cite{Aronov,ShelykhPRB72,Vasilopoulos07}} Despite its versatility, the approach relies on a free parameter ($\epsilon$ in Ref.[\textcolor[rgb]{0.00,0.00,1.00}{\onlinecite{Buttiker84}}]) used to characterize the QR-coupling with the leads, and modeled as a point-like scatterers. The free parameter $\epsilon$ represents the transmission amplitude  from one lead to the AB ring, or vice versa, and can take values in the range $0\le \epsilon\le 1/2$; where perfect reflection  means $\epsilon=0$, while perfect transmission entails $\epsilon=1/2$.
The introduction of this  parameter was a first effort to describe the possible band mismatch between the leads and the conducting AB-QR.
Furthermore the model introduces a second caveat, which arises when a third  scatterer is considered in one of the arms of the QR.  It predicts the counterintuitive result of a finite emerging current flux form the QR (non-zero electron transmission probability)
in the limit case of a zero dynamic phase change ($\theta=\pi R\sqrt{2m E_{\textsc f}/\hbar^2} $), {\it i.e.} eventhough the incident Fermi energy $E_{\textsc f}$ is set to zero and there is no bias voltage present.\textcolor[rgb]{0.00,0.00,1.00}{\cite{ShelykhPRB72,Vasilopoulos07}}
Clearly a special care have to be exercised when modeling the lead-to-ring for the case of semiconductor junctions.

Here we propose to replace the point-like scatterers that model the lead-to-ring junctions by two-dimensional quantum point contacts instead. By doing so, the dependence on a free parameter is naturally avoided in the model, and very important, the proposed geometry give us additional degrees of freedom to manipulate the conductance and spin interference in the QPC/QR$^{*}$/QPC device.

As we will show later, the conductance quantum oscillations in the QPC/QR$^{*}$/QPC device can be related not only to magnetic flux and the SOI effects as in earlier reports,\textcolor[rgb]{0.00,0.00,1.00}{\cite{ShelykhPRB72, Vasilopoulos07}} but also to the confinement strength of the  QPCs.  Indeed, this new feature (tunability of the QPCs) we have considered in the QR device allows means to control the spin-conductance oscillations of the output current, and in turn, can be used to modulate the Aharonov-Bohm and Aharanov-Casher phases solely by applying gate voltages.  Furthermore, it is predicted that the insertion of tunable QPCs in the QR leads to new and distinguishable physical response of the conductance oscillations of the QR, which are in principle, measurable at low temperatures. Hence a rich variety of interference phenomena arises which may be also of interest for potential applications.


\subsection{Quantum point contact model}
 \label{QPC-Model}

The lead-to-ring junctions of the device are assumed to form a narrow quantum constriction, {\it i.e.}, a QPC. Each QPC (including the one at the upper arm of the QR) are modeled through saddle point electrostatic potentials whose confinement are in principle tunable by external gates. To a good approximation the QPC can be represented through a purely quadratic two-dimensional potential,
\begin{equation}
 \label{SP-potential}
   V_{\mb{\tiny SP}}(x,y) = V_{o} - \frac{1}{2}m\,\omega_x^{2}\,x^{2} + \frac{1}{2}m\,\omega_y^2\,y^{2} .
\end{equation}
\noindent where $V_{o}$ is the potential energy at the saddle point, $m$ is the effective mass of the conduction electrons, and $\hbar\omega_{x}$ and $\hbar\omega_{y}$, are the characteristic longitudinal and the transversal energy confinements, respectively. The reference energy $V_{o}$ as well as the lateral $x$-$y$ confinement, can be adjusted in the actual experiments through electrical gates in an independent way. Remarkably, as deduced by Fertig and Halpering,\textcolor[rgb]{0.00,0.00,1.00}{\cite{Fertig87}} in the presence of a magnetic flux - due to a perpendicular magnetic field $B$ - the electron transmission amplitude of the two-dimensional potential (\ref{SP-potential})  can be described solely, by a dimensionless parameter $\varepsilon$, which is a function of all the relevant energies  characterizing the QPC. Namely $\varepsilon=\varepsilon(E_{\textsc f},\omega_{x},\omega_{y},\omega_{c})$, where
$\omega_{c} = eB/mc$ is the cyclotron frequency. The quantum transmission amplitude reads,\textcolor[rgb]{0.00,0.00,1.00}{\cite{Fertig87}}
\begin{equation}
\nonumber
t_{s}(\varepsilon)=\frac{1}{2}\left( \frac{ \Gamma(\frac{1}{4}-\frac{1}{2}i \varepsilon)}{\Gamma(\frac{1}{4}+\frac{1}{2} i \varepsilon)} e^{i\pi/4}-\frac{\Gamma( \frac{3}{4}-\frac{1}{2}i \varepsilon)}{\Gamma(\frac{3}{4}+\frac{1}{2} i \varepsilon)} e^{-i\pi/4} \right),
\end{equation}
\noindent where $\Gamma(z)$ is the Gamma function. The transmission probability  of the QPC, $T_{s} = |t_{s}|^2$ simplifies to
\begin{equation}
\label{T-Fertig}
  T_{s}(\varepsilon) = \frac{1}{1+ \text{exp}({-\pi \varepsilon})},
\end{equation}
\noindent which satisfies $T_{s}(\varepsilon) + R_{s}(\varepsilon) = 1$, and thus
\begin{equation}
\label{R-Fertig}
  R_{s}(\varepsilon) = \frac{1}{1+ \text{exp}({\pi \varepsilon})},
\end{equation}
\noindent will be the reflection probability that an incoming electron with unitary amplitude has been elastically backscatter from the QPC. In the limit $|\varepsilon|\gg 1$, we have for $\varepsilon>0$, $T_s(\varepsilon)\approx 1-e^{-\pi \varepsilon}$, while for $\varepsilon<0$, we have $T_s(\varepsilon)\approx e^{-\pi |\varepsilon|}$. Explicitly, the parameterized energy is given by $\varepsilon = (E_{G}-V_{o})/E_{1}$, and give us a measure of the energy of the semi-classical guiding-center motion $E_G$
relative to $V_{o}$. Here $E_{G} = E-(n+\frac{1}{2})E_2$,  with $n$ (non-negative integer) denoting the quantum index for the \emph{quasi}-Landau levels. The energies $E_\nu$ with $\nu=1,2$ are given explicitly by
\begin{equation}
 E_{\nu} = \frac{\nu\hbar}{2\sqrt{2}}
  \sqrt{\sqrt{\Omega^4+4\omega_x^2\omega_y^2}+(-1)^{\nu}\Omega^2}\,, \\
\end{equation}
\noindent with $\Omega^2=\omega_c^2+\omega_y^2-\omega_x^2$. Since in the limit of $\omega_{x,y} \ll \omega_{c}$, $E_{2}\approx \hbar\omega_c$, then $\Omega$ is called the effective oscillator frequency.\textcolor[rgb]{0.00,0.00,1.00}{\cite{Fertig87}}  Semi-classically, a charge carrier entering a region crossed by a magnetic field $B$, perpendicular to 2DEG and to the QR plane, will describe a circular motion with cyclotron frequency $\omega_{c}$. It is clear that the intensity of the scattering magnitudes (\ref{T-Fertig}) is directly bonded to $E_{G}$. Therefore, one can distinguish two different scenarios for the electronic transport, namely: ({\it i}) quantum tunneling for $E_{G} < V_{o}$, and ({\it ii}) quantum transmission following quasi-classical motion around the saddle-point potential for $E_{G} > V_{o}$. Here we will be interested in the latter case, with the cyclotron frequency written in terms of the magnetic flux penetrating the QR as $\omega_c=(e/\pi mc )\Phi/R^2$, where $\Phi = \pi R^2 B$ is the magnetic flux.

Before we go into the details of the scattering process in the QPC/QR$^{*}$/QPC device, it is convenient to revisit first the Hamiltonian model used to model the AB-AC-QR without scatterer (QPCs) and in presence of Rashba-SOI.

\section{AB-AC Quantum Ring Hamiltonian with SOI}
\label{QR-Hamiltonian}

The single-electron two-dimensional Hamiltonian for a semiconductor quantum ring in the presence of Rashba-SOI (without scatterers) and in the absence of a magnetic field is first  described in polar cylindrical coordinates. It reads
\begin{equation}
 \label{H2D}
 H(r,\varphi) = H_{o}(r) + H^{\prime}(r,\varphi),
\end{equation}
\noindent where the purely radial term is given by\textcolor[rgb]{0.00,0.00,1.00}{\cite{Meijer}}
\begin{equation}
 \label{Ho}
   {H_{o}(r)} = -\frac{\hbar^{2}}{2m} \left(\frac{\partial^{2}}{\partial r^{2}} + \frac{1}{r} \frac{\partial}{\partial r}\right ) + V(r),
\end{equation}
\noindent  whereas the angular dependence as well as the Rashba-SOI term, is contained in
\begin{equation}
 \label{H1}
   H^{\prime}(r,\varphi) = -\frac{\hbar^{2}}{2m\,r^{2}}\frac{\partial^2 }{\partial \varphi^2} -i\alpha_{so}\bn{\sigma}\cdot  \left( {\hat{\bn{e}}}_{r}\frac{1}{r}\frac{\partial}{\partial \varphi} - {\hat{\bn{e}}}_{\varphi} \frac{\partial}{\partial r} \right ),
\end{equation}
\noindent where $\alpha_{so}$ stands for the Rashba-SOI parameter strength, the matrix vector $\bn{\sigma}=({\sigma}_{x}, \sigma_y,\sigma_z)$, being $\sigma_{x,y,z}$, the usual spin Pauli matrices, ${\hat{\bn{e}}}_{r}=(\cos\varphi,\sin\varphi,0)$ and  ${\hat{\bn{e}}}_{\varphi} = (-\sin\varphi,\cos\varphi,0)$ are unitary vectors along the radial and azimuthal direction, respectively. Following Meijer {\it et al.},\textcolor[rgb]{0.00,0.00,1.00}{\cite{Meijer}} in the limit of a narrow ring ({\it i.e.} strong radial confinement), the associated confining energy in the radial direction becomes (typically) much larger than the characteristic SOI and kinetic energies of the circling electrons around the ring. Thus $H^{\prime}(r,\varphi)$ can be treated perturbatively by introducing an appropriate model potential $V(r)$ by solving first the purely radial term $H_{o}(r)$. An effective one-dimensional (1D) Hamiltonian depending only on the azimuthal angle $\varphi$ can be obtained  by taking the expectation value of Eq.\eqref{H1} with the eigenfunctions of the Hamiltonian $H_{o}$ in the lowest radial mode, $H_{1D}(\varphi)=\langle R_{o}(r) | H^{\prime}(r,\varphi)|R_{o}(r)\rangle$. The effective 1D Hamiltonian takes the form\textcolor[rgb]{0.00,0.00,1.00}{\cite{Meijer,Bulgakov02}}
\begin{equation}
 \label{1DHamilton}
   {H_{1D}} = \frac{\hbar^{2}}{2mR^{2}}\left\{ \left(i\frac{d}{d\varphi} + k_{so}R
    (\bn{\sigma}\cdot  {\hat{\bn{e}}}_{\varphi})\right)^{2} -(k_{so}R)^2\right\},
\end{equation}
\noindent with $k_{so}=m\alpha_{so}/\hbar^2$. Then to first order, we can rewrite (\ref{H2D}) as
\begin{equation}
 \label{H2Dsep}
 H(r,\varphi) = H_{o}(r) + H_{1D}(\varphi),
\end{equation}
\noindent as a separable in $r$ and $\varphi$ Hamiltonian to deal with.

\begin{figure}
\hspace*{-0.3cm}\includegraphics[width=3.5in,height=2.7in]{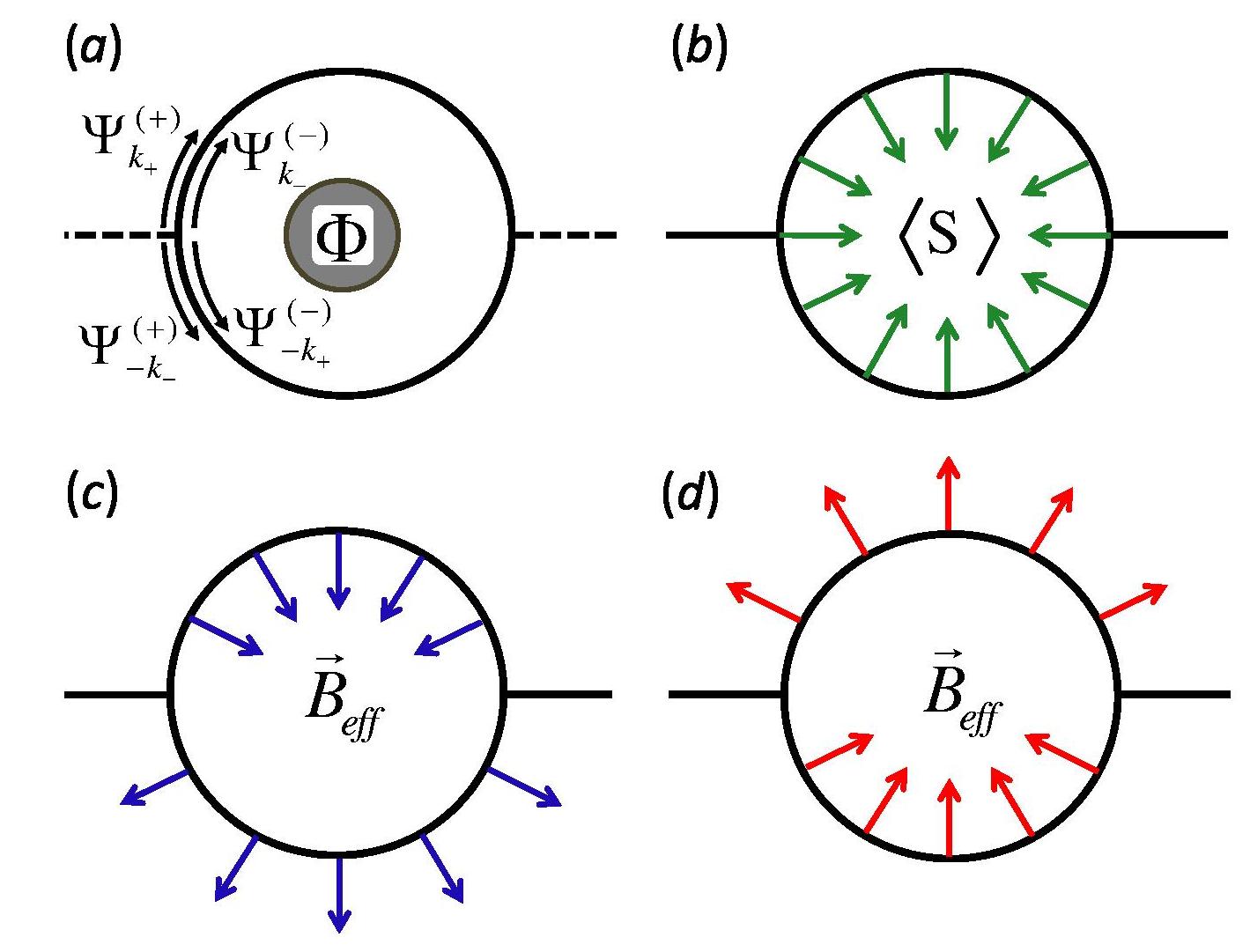}
\vspace{-0.1in}
 \caption{\label{EigenWave} (Color online) (a) The diagram sketches the clockwise and anti-clockwise  propagation for the two spin-resolved eigenstates, defined in (\ref{Eigenstates-plus}) and (\ref{Eigenstates-minus}) within the adiabatic approximation. (b) Spin orientation $\langle \bn S \rangle$ of the electrons winding the ring. (c) and (d) Rashba induced effective magnetic field $\bn{B}_{eff}$ for clockwise and anti-clockwise cycling electrons, respectively.}
 \end{figure}

Up to this point we have not considered yet the effect of an external magnetic field (magnetic flux) across the path of the winding electrons in the ring. This is done by introducing the usual {\it minimal coupling} substitution of the momentum operator, $-i\nabla\rightarrow -i\nabla -e\bn{A}$. The vector potential can be conveniently chosen to be tangential to the QR for a magnetic field $B$ perpendicular to the plane of the ring by using ${\bn A}= (BR/2){\hat{\bn{e}}}_{\varphi}$. As a result the moving electrons flowing around the ring will experiment a magnetic flux given by $\Phi = \oint {\bn A}\cdot(R\,d\varphi\, {\hat{\bn{e}}}_{\varphi})= \pi R^2B $. This is formally equivalent to the replacement $id/d\varphi \rightarrow id/d\varphi + \Phi/\Phi_{o}$ in Eq.\eqref{1DHamilton}, where $\Phi_{o} = h/e$ is the the magnetic flux quanta. The latter gives rise to the well known AB phase ($\Phi_{AB}=2\pi\Phi/\Phi_o$) acquired by the electrons moving in the close path of the quantum  in the presence of a magnetic flux. As a consequence the physical information of AB phase can be carried out directly in the overall phase of the eigenstates of the SOI 1D-Hamiltonian \eqref{1DHamilton}.\textcolor[rgb]{0.00,0.00,1.00}{\cite{Bulgakov02,Shelykh05-2}} Note that the magnetic field may or not penetrate the QR. Here we are interested in the relatively weak magnetic field regime, or either the case in which just the magnetic flux penetrates the AB-AC-QR (see Fig.\ref{EigenWave}a), but not the field itself, and therefore the Zeeman effect can be safely neglected in the full description.

Given that the projection of the total angular momentum along the $z$-axis commutes with the 1D Hamiltonian, $[J_z,H_{1D}]=0$, with $J_{z}= -i\hbar (\partial/\partial\varphi)-\hbar\sigma_z/2$. Then it follows that a particular set of eigensolutions of the system $H_{1D}|\Psi(\varphi)\rangle=E|\Psi(\varphi)\rangle$ per spin orientation and clockwise/anticlockwise $(+/-)$ propagation of the electrons can be written as\textcolor[rgb]{0.00,0.00,1.00}{\cite{ShelykhPRB72,ShelykhPRB71}}
\begin{equation}
 \label{Eigenstates-plus}
  |\Psi_{k_{+}}^{(+)}\rangle =
   e^{ik_{+}R\,\varphi}
    \,|+\rangle ,\;\;\;
  |\Psi_{k_{-}}^{(+)}\rangle =
   e^{-i k_{-}R\,\varphi}
     \,|+\rangle ,
\end{equation}
\begin{equation}
\label{Eigenstates-minus}
  |\Psi_{k_{-}}^{(-)}\rangle =
  e^{ik_{-}R\,\varphi}
    \,|-\rangle ,\;\;\;
  |\Psi_{k_{+}}^{(-)}\rangle =
   e^{-i k_{+}R\,\varphi}
    \,|-\rangle ,
\end{equation}
\noindent with normalized eigenspinors given by\textcolor[rgb]{0.00,0.00,1.00}{\cite{ShelykhPRB72}}
\begin{equation}
  |+ \rangle =
    \frac{1}{\sqrt{1+\eta^2}}\left[
    \begin{array}{c}
      i\eta \\
      e^{-i\varphi}
    \end{array}
    \right], \;\;
  |-\rangle =
     \frac{1}{\sqrt{1+\eta^2}}\left[
    \begin{array}{c}
      1 \\
      i\eta\, e^{-i\varphi}
    \end{array}
    \right],
\end{equation}
\noindent where the SOI-dependent dimensionless $\eta$ parameter is given by
\begin{equation}
 \label{spinor-coef}
 \eta = \frac{2k_{so}R}{1+\sqrt{1 + (2k_{so}R)^{2}}}.
\end{equation}

The corresponding spin-dependent wave numbers $k_{\sigma}$, with $\sigma = \pm$ satisfies the quadratic equation,
\begin{equation}
 \label{WavNum-Eq}
  \left(k_F^2 - k^{2}_{\sigma}\right)\left[k_F^2 -
  \left(k_{\sigma} - \frac{1}{R} \right)^{2}\right] - 4k_{so}^2
  \left(k_{\sigma} - \frac{1}{2R} \right)^{2} = 0.
\end{equation}
\noindent with $k_{\textsc f} = \sqrt{2mE_{\textsc f}/\hbar^2}$ the Fermi wave number.  For a sufficient large SOI and large ring radius to fulfill  the adiabatic approximation $k_{\pm} \gg 1/R$, the spin-resolved wave numbers solutions of Eq.(\ref{WavNum-Eq}) are recasted as,
\begin{equation}
 \label{WavNum-Val}
 k_{\sigma} = {\sigma}k_{so}  + \sqrt{k_F^2+k_{so}^2}\simeq k_F+{\sigma}k_{so}.
\end{equation}
\noindent the right hand side holds as long $k^2_{so}\ll k^2_F$, which is true for typical 2DEG embedded in III-V semiconductor quantum wells.

Let us now consider a full cycle of the wave functions around the QR under a penetrating magnetic flux. Apart from the usual Aharonov-Bohm phase $\Phi_{AB}$ discussed above due to the presence of the SOI, a topological Berry phase $\theta_B^{\pm}$ have to be added to the spin-dependent wave functions.
The Berry phase  arises as a consequence of the phase factor acquired by the electron spin at the end of its loop around the QR as it adiabatically rotates (precesses) trying to follow the effective magnetic field $ {\bn{B}}_{eff}$ due to SOI (see Fig.\ref{EigenWave}c,\ref{EigenWave}d).   That is, $|\Psi^{(+)}_{k_{\sigma}}\rangle\rightarrow e^{i \Phi_{AB}}e^{-i\theta_B^+}|\Psi^{(+)}_{k_{\sigma}}\rangle$ and $|\Psi^{(-)}_{k_{\sigma}}\rangle\rightarrow e^{i \Phi_{AB}}e^{-i\theta_B^-}|\Psi^{(-)}_{k_{\sigma}}\rangle$, and we have followed the sign convention in the Berry phase of Ref.\textcolor[rgb]{0.00,0.00,1.00}{\onlinecite{Aronov}}. The Berry phase is calculated as half of the material angle subtended by the effective magnetic field, such that the angle $\theta^{\pm}_B$ between the $z$-direction and the cone formed by the precessing spin satisfies
\begin{eqnarray}
\cot\theta_B^{\pm} & = &\frac{\langle\Psi^{(\pm)}_{k_{\sigma}} |{\sigma_z}|\Psi^{(\pm)}_{k_{\sigma}}\rangle}{\sqrt{ \langle\Psi^{(\pm)}_{k_{\sigma}} |{ \sigma_x}|\Psi^{(\pm)}_{k_{\sigma}}\rangle^2 +\langle\Psi^{(\pm)}_{k_{\sigma}} |{ \sigma_y}|\Psi^{(\pm)}_{k_{\sigma}}\rangle^2}} \nonumber\\
 & = & \pm\frac {\eta^2-1}{2\eta}=\mp \frac{\hbar^2}{2m\,\alpha_{so} R}
\end{eqnarray}

In the adiabatic approximation   $k_{so}R\gg 1$ ({\it i.e.} in the limit of vanishing spin-geometric phase\textcolor[rgb]{0.00,0.00,1.00}{\cite{Nitta12,KlausR}}) the parameter $\eta\rightarrow 1$ and the spin orientation of the electrons  is described by the vector $\langle\Psi^{(-)}_{k_{\sigma}} |{\bn \sigma}|\Psi^{(-)}_{k_{\sigma}}\rangle=-\langle\Psi^{(+)}_{ k_{\sigma}} |{\bn \sigma}|\Psi^{(+)}_{k_{\sigma}}\rangle=( \sin\varphi,\cos\varphi,0)$. That is, the electron spin is oriented in-plane, radially toward or from the ring center of the QR, see Fig.\ref{EigenWave}(b). Since the  Rashba spin-orbit induced effective magnetic field is given by $ \bn{B}_{eff} \sim \alpha_{so} (\bn{k} \times \bn{F}$), with $\bn{k}$ the wave vector of the moving electrons in an external electric field $\bn{F}$, then the spin-orientation is directed either, parallel or antiparallel to the effective magnetic field. For this case the overall Berry phase is just a constant and reduces to $\theta_B^{\pm}=\pm \pi/2.$

Clearly these results above holds as long there are no scatterer points along the path of the electrons in the QR. The role of the QPCs on the scattering process, together with the phases involved are the focus of the following discussion.

\section{Scattering Formalism}
\label{Sec:Scattering}

\begin{figure}
 \hspace*{-0.2cm}\includegraphics[width=3.3in,height=2.0in]{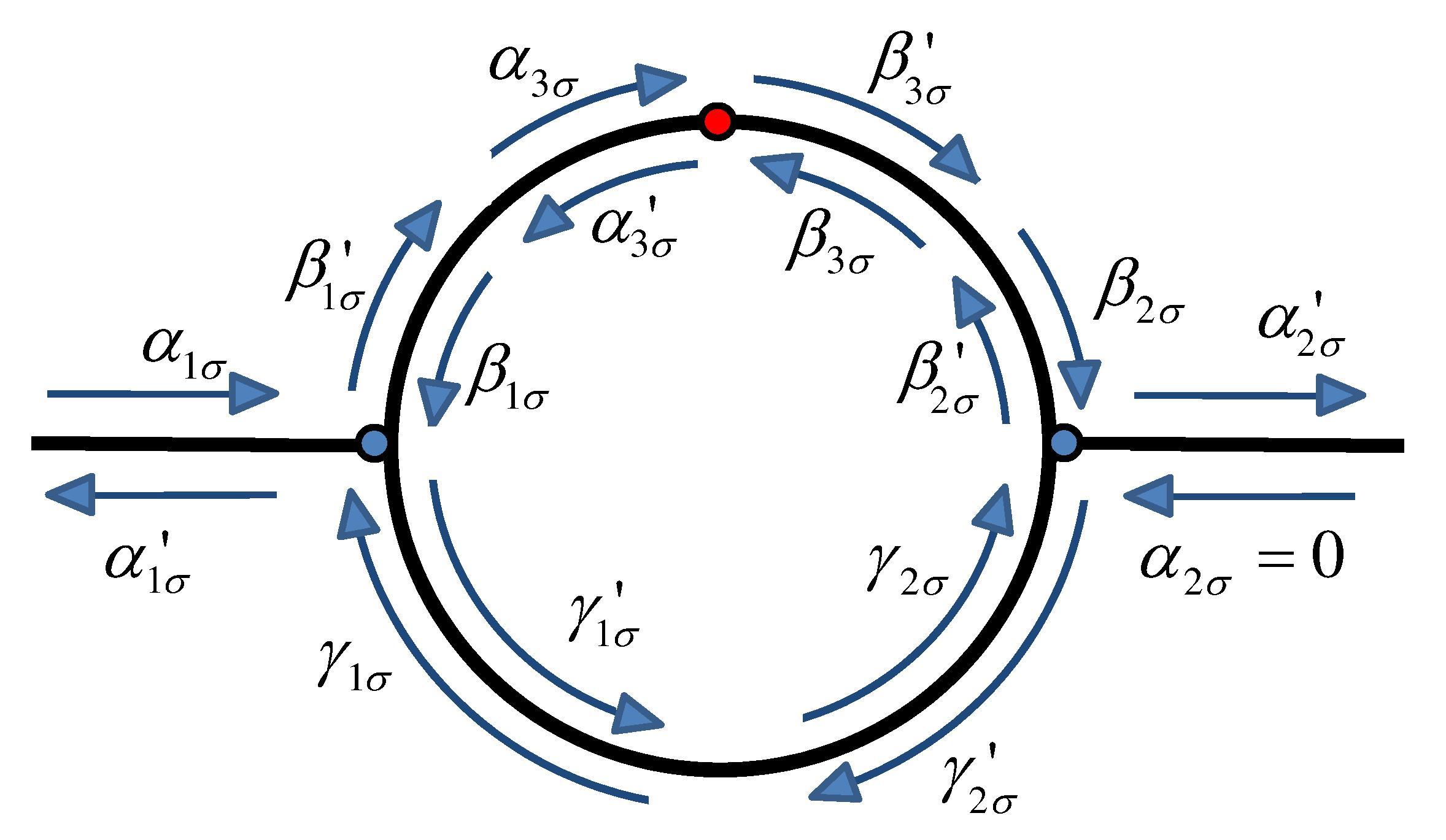}
 \vspace{-0.1in}
 \caption{\label{AmpliWave} (Color online) Diagram of the QR device attached to the leads. The arrows indicates the spin-resolved wave amplitudes during the spin-dependent scattering processes. The QPCs  are indicated schematically by the small blue and red circles. }
\end{figure}

As depicted in Fig.\ref{QR-device} at the junction of each lead with the QR is a QPC$_{i}$, where $i=1(2)$ refers to the left(right) quantum point contact. Additionally a third scatterer (QPC$_{3}$) is located at the upper arm of the QR.  Now, it is important to mention that spin scattering mechanisms such as the spin-orbit mediated coupling with (piezoelectric) phonons, are believed to be the main source of spin-fliping in QRs as well as in quantum dots. However it is also well known theoretically and experimentally, that the associated relaxation times in these systems are extremelly large,  ranging from milliseconds up to few seconds, see Refs.[37],[38] and [39].  This times are longer than the associated momentum and collision  times in such ballistic systems and will not play a role here.  Therefore any spin-flip processes at the QPCs will be subsequently neglected in this work. Also, without loss of generality, it will be assumed that the electron spin is preserved along the QR path.
The spin states can be treated then as two independent channels of propagation. Consequently, for each spin channel we have three outgoing electronic waves with amplitudes $(\alpha_{i\sigma}^{\prime},\beta_{i\sigma}^{\prime},\gamma_{i\sigma}^{\prime})$, and accordingly,  three incoming waves with amplitudes $(\alpha_{i\sigma},\beta_{i\sigma},\gamma_{i\sigma})$. Similarly, for the QPC$_{3}$  we have two ingoing and two outgoing electronic waves with amplitudes  $(\alpha_{3\sigma},\beta_{3\sigma})$ and  $(\alpha_{3\sigma}^{\prime},\beta_{3\sigma}^{\prime})$, respectively (see Fig.\ref{AmpliWave}).

It is worth recalling that in the adiabatic approximation, the presence of the Rashba-SOI in the QR do not actually intermixes the electron spins. The Rashba-SOI enters here solely a source of an AC phase of the traveling waves, as has been previously discussed. The AC phase adds up then to the AB and Berry phase acquired by the electrons waves in their round trip along the QR. We chose the convention that clockwise-oriented waves give rise to positive phases, whereas the waves moving in the opposite direction accumulates a negative phase. 

Each lead-to-ring junction (the scattering of an electron with spin $\sigma$ at QPC$_{1,2}$) is described by a ($3\times3$) scattering matrix $\bn{\mathbb{S}}$. Under the presence of a magnetic flux $\Phi$ penetrating the QR, time reversal invariance (TRI) is satisfied, leading to the known Onsager-Casimir relations for the scattering matrix elements $S_{ij}(\Phi) = S_{ji}(-\Phi)$, where $i,j$ are the outgoing/ingoing channel, respectively. In the absence of the magnetic flux it reduces to the expected reciprocal relations (symmetricity of $\bn{\mathbb{S}}$). This physically implies that the symmetric injection-ejection process of electrons at each junction (QPC) of the QR device is withhold. In addition, unitarity of $\bn{\mathbb{S}}$ ensures current conservation at both, left and right QPCs. Therefore we can write
\begin{equation}
 \label{1-SM-2}
 \begin{bmatrix}
  \alpha_{i\sigma}^{\prime}\\\\
  \beta_{i\sigma}^{\prime}\\\\
  \gamma_{i\sigma}^{\prime}
 \end{bmatrix} = \bn{\mathbb{S}}
 \begin{bmatrix}
  \alpha_{i\sigma}\\\\
  \beta_{i\sigma}\\\\
  \gamma_{i\sigma}
 \end{bmatrix}, \;\;\;\mb{with $i=1,2$},
\end{equation}
\noindent as required for outgoing and ingoing waves amplitudes from standard scattering formalism. Given that flux conservation together with TRI, guarantee $\bn{\mathbb{S}}^{-1}=\bn{\mathbb{S}}^{*}$ which entails the symmetricity property $\bn{\mathbb{S}}$, allow thus to propose
\begin{equation}
 \label{SM12-sim}
  \bn{\mathbb{S}} = \left( \begin{array}{ccc}
     R_{s}^{1/2} & \textsf{a}T_{s}^{1/2} & \textsf{a}T_{s}^{1/2}\\
     \textsf{a}T_{s}^{1/2} & \textsf{b} & \textsf{c} \\
     \textsf{a}T_{s}^{1/2} & \textsf{c} & \textsf{b}
  \end{array} \right),
\end{equation}

\noindent where we have defined $T_{s}$, as the electron transmission amplitude with spin $\sigma$ through a given QPC after formula (\ref{T-Fertig}),  fulfilling $T_{s}(\varepsilon) + R_{s}(\varepsilon) = 1$. The parameters $\textsf{a}, \textsf{b}$ and $\textsf{c}$  are real coefficients to be determined. In (\ref{SM12-sim}) the diagonal matrix elements, $S_{ii}$, represent the reflection amplitude of the $i-th$ channel, whereas the off-diagonal elements $S_{ij}$ ($i\ne j$) characterizes the transmission amplitudes from channel $i$ to $j$ at a given junction.  By demanding the unitary condition of the incident flux, {\it i.e.} $\bn{\mathbb{S}}^{\dag}\cdot\bn{\mathbb{S}} = \bn{I}_{3}$, with $\bn{I}_{N}$ the \textbf{($N \times N$)} identity matrix, the  parameters $\textsf{a}, \textsf{b}$ and $\textsf{c}$ in (\ref{SM12-sim}) are readily obtained, yielding
\begin{equation}
 \label{S-coeff}
 \textsf{a} = \frac{1}{\sqrt{2}}; \;\; \textsf{b} = \frac{-R_{s}^{1/2} \pm 1}{2};
 \;\,\mb{and} \;\; \textsf{c} = \frac{-R_{s}^{1/2} \mp 1}{2},
\end{equation}
\noindent notice that \textsf{b} and \textsf{c} depend explicitly on the effective energy $\varepsilon$ through the reflection probability of each QPC.

Let us consider now the incident and reflected electronic waves at one fourth cycle in the upper branch of the QR, just at the QPC$_{3}$ location (see Fig. \ref{AmpliWave}). In such a case, by following usual rules of transfer matrix formalism,\textcolor[rgb]{0.00,0.00,1.00}{\cite{Buttiker84}} the amplitudes of the wave functions  can be related to those of the QPC$_{1}$ as follows
\begin{eqnarray}
 \label{tra-ref(31)}
     \left[
    \begin{array}{l}
      \alpha_{3\sigma} \\
      \alpha\p_{3\sigma}
    \end{array}
  \right] & = &
  \left(
   \ba{cc}
      \text{exp}\,\left(i \varphi_{2\sigma}\right) & 0 \\
      0 &  \text{exp}\,\left(-i \varphi_{1\sigma}\right)
   \ea
  \right)
   \left[
    \begin{array}{l}
      \beta\p_{1\sigma} \\
      \beta_{1\sigma}
    \end{array}
  \right]  ,
\end{eqnarray}
\noindent and similarly with respect to the wave amplitudes at   QPC$_{2}$,
\begin{eqnarray}
 \label{tra-ref(32)}
   \left[
    \begin{array}{l}
      \beta_{2\sigma} \\
      \beta\p_{2\sigma}
    \end{array}
  \right] & = &
   \left(
   \ba{cc}
     \text{exp}\,\left(i \varphi_{2\sigma}\right) & 0 \\
      0 &  \text{exp}\,\left(-i \varphi_{1\sigma}\right)
   \ea
  \right)
   \left[
    \begin{array}{l}
      \beta\p_{3\sigma} \\
      \beta_{3\sigma}
    \end{array}
   \right],
 \end{eqnarray}
\noindent where $\varphi_{1}$ is the net phase acquired by an electron traversing half of the upper arm semicircle in the clockwise direction, whiles $\varphi_{2}$ corresponds to the phase acquired by the electron in the same segment but in the counterclockwise direction. For spin up/down ($\sigma=\pm$) electrons such phases read explicitly
\begin{equation}
 \label{Phase1}
 \varphi_{1\pm}=\frac{\pi}{2}(k_{\pm}R - \phi + \frac{1}{2}),
\end{equation}
\begin{equation}
 \varphi_{2\pm}= \frac{\pi}{2}(-k_{\mp}R - \phi + \frac{1}{2}),
\end{equation}
\noindent with $k_\sigma$ as given by of Eq.(\ref{WavNum-Val})  and $\phi =\Phi/\Phi_{o}$ is the magnetic flux in units of the magnetic flux quanta ($\Phi_{o}=h/e$). The added quantity $\pi/4$ comes from the accumulated Berry phase in the semicircle. Note that due to the Rashba-SOI, Kramers degeneracy is broken, and hence an electron --at a given Fermi energy--, with spin-up orientation $(\sigma= +)$ and traveling with a wavenumber $k_{+}$ will be elastically backscattered with a wavenumber $-k_{-}$, as given by Eq. (15). Similarly occurs for the opposite spin case $(\sigma=-)$.

For the lower arm of the QR the wave amplitudes are transferred according to
\begin{eqnarray}
\label{tra-ref(Lw)}
     \left[
    \begin{array}{l}
      \gamma_{1\sigma} \\
      \gamma\p_{1\sigma}
    \end{array}
  \right] & = &
  \left(
   \ba{cc}
      \text{exp}\,\left(i \varphi_{4\sigma}\right) & 0 \\
      0 &  \text{exp}\,\left(-i \varphi_{3\sigma}\right)
   \ea
  \right) \mathbb{T}_{l}
   \left[
    \begin{array}{l}
      \gamma\p_{2\sigma} \\
      \gamma_{2\sigma}
    \end{array}
  \right],
\end{eqnarray}
\noindent with $\varphi_{3\sigma} = 2\varphi_{1\sigma}$ and $\varphi_{4\sigma} = 2\varphi_{2\sigma}$. In equation (\ref{tra-ref(Lw)}) we have included for completeness the presence of a fourth scatterer (QPC$_4$) at the lower arm of the QR, characterized by the transfer matrix $\mathbb{T}_{l}$. For the case in which a constriction-free path in the lower branch of the QR is considered, then we use  $\mathbb{T}_{l} = \bn{I}_{2}$.

After some standard scattering theory operations within the transfer matrix approach, together with TRI symmetry properties lead us to express
\begin{equation}
 \label{tra-ref(3)}
    \left[
    \begin{array}{l}
      \beta\p_{3\sigma} \\
      \beta_{3\sigma}
    \end{array}
  \right] =
 \mathbb{T}_{u}
 \left[
    \begin{array}{l}
      \alpha_{3\sigma} \\
      \alpha\p_{3\sigma}
    \end{array}
  \right],
\end{equation}
\noindent in which the transfer matrix $\mathbb{T}_{u}$ has the form
\begin{equation}
 \mathbb{T}_{u}=\left(
   \ba{cc}
     \frac{1}{t_s} & -\frac{r_s}{t_s} \\
     -\frac{r^{*}_s}{t^{*}_s} & \frac{1}{t^{*}_s}
   \ea
 \right),
\end{equation}
\noindent and we will assume that $t_{s}(\varepsilon) = T_{s}(\varepsilon)^{1/2}$ and $r_{s}(\varepsilon) = R_{s}(\varepsilon)^{1/2}$. They stand for the transmission and reflection amplitudes of the QPC$_{3}$, respectively. On the other hand, it is convenient to define the diagonal matrix
\begin{equation}
 \mathbf{\Phi}^{\sigma}_{u,l} =
  \left(
   \ba{cc}
      \text{exp}\,\left(i \varphi_{2\sigma,4\sigma}\right) & 0 \\
      0 & \text{exp}\,\left(-i \varphi_{1\sigma,3\sigma}\right)
   \ea
  \right),
\end{equation}
\noindent where the subscripts $(u,l)$ stands for $(upper,\,lower)$ part of the QR, respectively. After some algebraic manipulations (see appendix A), a closed expression is obtained for the total transmitted amplitude $\alpha\p_{2\sigma}$ of the outgoing electrons at the right lead of the QR, yielding
\begin{equation}
 \label{AmpTra-QR}
   \left[
    \begin{array}{c}
      \alpha\p_{2\sigma} \\
      0
    \end{array}
  \right]  =
 \frac{2 T_{s}}{\left( R_{s}^{1/2}+\lambda \right)^2}
    {\bn \lambda}\,(\mathbb{P}^{\,\sigma})^{-1}\mathbf{\Phi}^{\sigma}_{u}\mathbb{T}_{u}
   \mathbf{\Phi}^{\sigma}_{u}
    \left[
       \begin{array}{c}
         -\lambda \\
         1
       \end{array}
    \right],
\end{equation}
\noindent  where $\mathbb{P}^{\,\sigma}$ is a $2\times2$ matrix depending upon the QR parameters (Appendix A) while the matrix $\bn \lambda$ is given by
\begin{equation}
 {\bn \lambda}=\left(
    \begin{array}{cc}
      \lambda & 1\\
      0 & 0
       \end{array}
  \right); \, \lambda=\pm 1.
\end{equation}

It is worth to emphasize that Eq.(28) summarizes one of the main results of this work. Its importance relies on the fact that, all the relevant physical information of the coherent transport phenomena, such as the electron scattering, the change of phase (including AB, AC and Berry phases) as well as the interference phenomenon occurring in the QR setup, are all comprised in $\alpha\p_{2\sigma}$.
The two-probe Landauer conductance,\textcolor[rgb]{0.00,0.00,1.00}{\cite{Landaue89}} of the QR is described by
\begin{equation}
 \label{G-Ncompo}
  G(\varepsilon) = \sum_{\sigma} G_{\sigma}  (\varepsilon) = \frac{e^{2}}{h}\sum_{\sigma}\left|\alpha\p_{2\sigma}(\varepsilon)\right|^{2},
\end{equation}
\noindent then it is possible to find a closed analylical formula for the spin-resolved conductance of the QR device, and  it is given by
\begin{widetext}
\begin{equation}
 \label{G-spinres}
  G_{\sigma}(\varepsilon)= \frac{e^{2}}{h}\frac{4T_{o}^2(\varepsilon)[1+T_{s}(\varepsilon) +2\sqrt{T_s(\varepsilon)}\cos(2\varphi_{\sigma}) ]{\sin}^2\theta}
  {[ A_{-}(\varepsilon)-A_o(\varepsilon)\cos(2\theta)+A_{+}(\varepsilon)\sqrt{T_s(\varepsilon)}
  \cos(2\varphi_{\sigma})]^2+4T_o^2(\varepsilon)\sin^2(2\theta)},
\end{equation}
\end{widetext}
\noindent where we have defined $A_{\pm}(\varepsilon)= 2-T_{o}(\varepsilon)\pm 2\sqrt{1-T_o(\varepsilon)}$ and $A_o(\varepsilon)=A_{+}(\varepsilon)+A_{-}(\varepsilon)$. The angular parameters describing the different phases are namely, the dynamical phase $\theta=\pi k_{\textsc f}R$, and aspin-dependent phase
\begin{equation}
\varphi_{\sigma} = \pi\left(\frac{1}{2}- \phi + \sigma \phi_{so}\right)
\end{equation}
\noindent that contain all the involved phases, with  $\phi$ and $\phi_{so} = k_{so}R$ being the AB and the AC phases, the $\pi/2$ comes from the Berry phase. The transmission coefficient is the same in both QPC$_{1,2}$, $T_{o}(\varepsilon)$, as they are supposed to be identical by construction, while the same quantity at QPC$_{3}$ is denoted by $T_{s}(\varepsilon)$. We emphasize though that the formalism is quite general and admits different transmission probabilities of QPC$_{1}$ and QPC$_{2}$.  Here for the sake of clearity, we have considered only the case in which both QPCs are identical. This allow us to concentrate on the interplay of other important parameters such varying the aperture of QPC$_{3}$ whiles the magnetic field and/or the Rashba spin-orbit coupling are tuned. As shown above by considering symmetric QPCs leads us to arrive to a relatively simple and explicit formula for the spin conductance of the device (Eq. 31) which in turn enable us to a deeper understanding of the resonance and anti-resonance behavior of the QR device, discussed in further detail below.

Note that the conductance result of Eq.(\ref{G-spinres}) hold in general, that is, independent of the transmission probability model $T_{s}$ considered for the QPCs. Here we will use the saddle point potential profile of Fertig and Halpering discussed in Sec.II to model the transmission probability of the QPCs, explicitly Eq.(\ref{T-Fertig}) and Eq.(3). Due to its strong dependence on the magnetic flux (magnetic field), as well as with the lateral and transversal electrostatic confinement at the constrictions, the QR device proposed  offers additional degrees of freedom to manipulate  spin-transport and interference phenomena.  Indeed, as predicted by Eq.(\ref{G-spinres}), the particular geometry choice of the QR device with tunable QPCs, results in a conductance which exhibits strong oscillations with the strength of the magnetic flux and Rashba-SOI, oscillations which in turn are modulated in a non-trivial way by the opacity of the QPCs, as it will be discussed later in more detail.

The cases with different confinement strengths for  QPC$_{1}$ and QPC$_{2}$, together with the one of a fourth QPC$_{4}$ at the lower arm of the QR  are beyond the scope of the present work and will be treated in a subsequent publication. We anticipate though that due the cumbersome expressions in such cases, it would render quite difficult to arrive to close analytical formulas of the conductance of the QR device for such cases.

\subsection{ Spin resolved conductance: limit cases }
\label{Sec:Limits}

It is illustrative to analyze the spin resolved conductance at some limit cases. We can distinguish three extreme scenarios for the conductance behavior depending upon the opacity or the aperture of the QPCs. Namely, case ({\it i}), maximum transparency at each QPC$_{i}$, case ({\it ii}), maximum transparency only at QPC$_{3}$, and case ($iii$) maximum transparency at QPC$_1$ and QPC$_2$ whereas  QPC$_3$ exhibits a variable transparency.

For {\it case ({\it i})} we have $T_o(\varepsilon)= T_{s}(\varepsilon) = 1$, thus $A_{\pm}=1$, $A_{o}=2$, and  the spin-resolved (\ref{G-spinres}) conductance is readily compacted to
\begin{equation}
\label{G-QRFree}
 G_{\sigma}= \frac{e^{2}}{h}\left[ \frac{16\cos^2\varphi_{\sigma} {\sin}^2\theta}
 {\left(1-2\cos(2\theta)+\cos(2\varphi_{\sigma})\right)^2+4\sin^2(2\theta)}\right ],
\end{equation}
\noindent which is only a function of the dynamic phase $\theta$, and the round trip  phase $\varphi_{\sigma}$ due to the accumulated AB, AC and Berry phases. The expression between large square brackets, is just the spin-dependent transmission probability of the QR. In the absence of Rashba-SOI and  neglecting the Berry phase ({\it i.e.} with $\varphi_{\sigma}  =  - \pi\phi$) it reduces to the formula derived by Vasilopoulus {\it et al.}\textcolor[rgb]{0.00,0.00,1.00} {\cite{Vasilopoulos07}} for the transmission probability in an AB-QR without scatterers [Eq.(7) in Ref.\textcolor[rgb]{0.00,0.00,1.00}{\onlinecite{Vasilopoulos07}}]. Clearly formula (\ref{G-QRFree}) leads to resonances and antiresonances in the total conductance. The condition for the appearance of resonances with a maximum in the spin-conductance ($G_{\sigma} = 1$) are hold whenever $\varphi_{\sigma}= n\pi$, with $n$ integer. On the other hand, all the antiresonances (zeros of the spin-conductance) appear when $\varphi_{\sigma}=\frac{\pi}{2}n$. By fixing the magnetic flux $\phi=\frac{1}{2}$ then these maximum resonances/anti resonances are present as long $\phi_{\textsc {so}}^{max} = n$ and $\phi_{\textsc {so}}^{min} = \frac{n}{2}$, with $n$ an integer and odd number, respectively.

{\it Case ({\it ii})}: perfect transparency at QPC$_{3}$ only, {\it i.e.}  backscattering-free regime in QPC$_{3}$, which entails $T_{s}(\varepsilon) = 1$ together with $T_{o}(\varepsilon) \neq 1$ in the other two QPCs. Considering these requirements the condition for maximum conductance is dictated by
\begin{widetext}
 \begin{equation}
  \label{G-max2}
  \text{cos}(2\phi_{\sigma}) = \frac{T_{o}(\varepsilon)^2 + 4\left[2\left(1+\sqrt{R_{o}(\varepsilon)}\right)-
  \left(2+\sqrt{R_{o}(\varepsilon)}\right)
  T_{o}(\varepsilon)\right]\cos(2\theta)}
  {\left(T_{o}(\varepsilon)-2-2\sqrt{R_{o}(\varepsilon)}\right)^2}.
 \end{equation}
\end{widetext}

{\it Case ({\it iii})}: perfect transparency at QPC$_{1}$ and QPC$_{2}$ ($T_{o}(\varepsilon) = 1$) whereas $T_{s}(\varepsilon) \neq 1$ in QPC$_{3}$.  Here the maximum spin-conductance is expected as long that
\begin{equation}
 \label{G-max3}
 \text{cos}(2\phi_{\sigma}) = \frac{ 1\pm \sqrt{R_{s}(\varepsilon)}\sin\theta  }{\sqrt {T_{s}(\varepsilon)}}.
\end{equation}

These expressions above will be useful in the understanding  of the origin of the QR-conductance oscillations and will be discussed in more detail in the next section. Finally, we should mention that there is yet another possible configuration. This is such that QPC$_{3}$ is totally opaque (closed) with $R_{s}(\varepsilon) = 1$ ({\it i.e.} $T_{s}(\varepsilon) = 0$) and finite transmitivity at the other two QPCs. From Eq.(\ref{G-spinres}) we can infer that the AB and AC oscillations are totally absent in this situation, as they should, due the truncated path for the moving electrons in the upper arm of the QR because of the suppressed quantum interference.

\section{Numerical Results and Discussion}
 \label{Sec:Results}

\begin{figure}
 \hspace*{0.0in}
 \includegraphics[width=3.8in,height=4.7in]{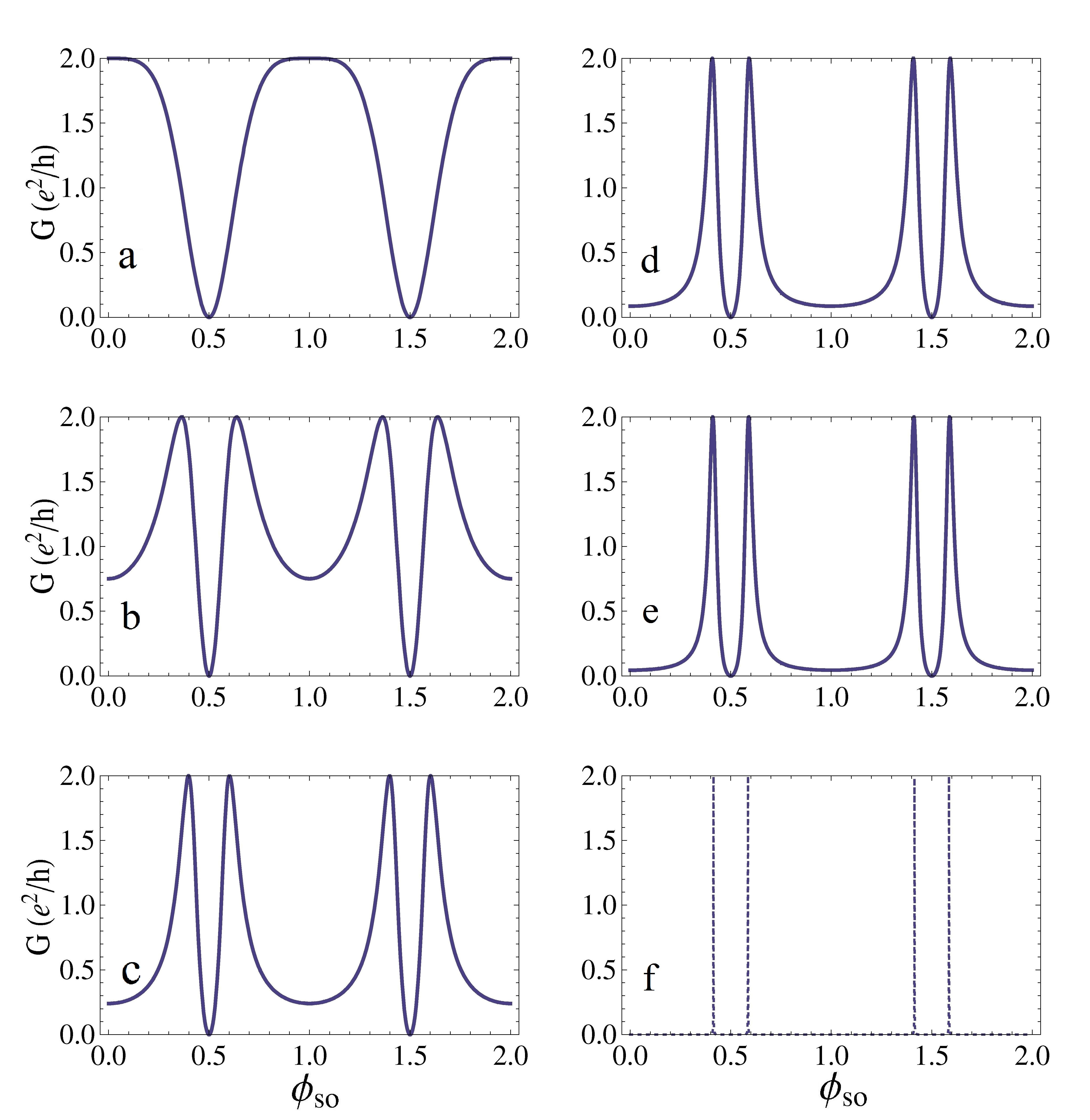}
\vspace{-0.1in}
  \caption{\label{F1-QR} Evolution of the conductance quantum oscillations as a function of the AC phase $\phi_{\textsc {so}}$. We have fixed $T_{s}(\varepsilon) = 1$   and the AB-phase to $\phi = 0.5$, the latter corresponds to a field of $B=10$ mT for the QR structure considered here. The transmission coefficients used at the lead-to-ring junctions, in the panels {\it a, b, c, d, e} and {\it f}; correspond to  $T_{o}(\varepsilon) = 1, 3/4, 1/2, 1/3, 1/4$ and $10^{-3}$, respectively, or equivalently, to confinement energies strenghts of $\hbar \omega_j=0.2, 2.44, 2.72, 2.94, 3.08$ and $10.2$ meV.}
\end{figure}

Unless otherwise specified, the numerical results reported here were calculated using a QR radius of $R = 250$ nm defined on an $InAs$-based heterostructure with electron effective mass of $m = 0.023$. Hereafter, the two-probe Landauer conductance $G(\varepsilon)$ (\ref{G-Ncompo}) will be referred   as $G$, whiles  for simplicity the characteristic longitudinal and transversal energy confinements at the \textit{lead-to-ring} junction  will be assumed to be equal ($\hbar\omega_{x} = \hbar\omega_{y}$), and from now on denoted by $\hbar\omega_{j}$. Notice that  even large changes between them will not change the main physical results as the transmission probabilities defined in Eq.(2) will follow essentially the same step-like trend in such cases.\cite{MButtiker} The typical values for  $\hbar\omega_{j}$ are in the range of $0.8\sim 3$ meV\cite{vanHouten,Gilbert}, whereas the experimental values of the Rashba parameter $\alpha_{\textsc {so}}$  are between  $20\sim 40$ meV\,nm (in InAs-based 2DEGs \cite{Sato,Nitta03}) and will correspond to  AC-phases  $\phi_{\textsc {so}}$ of about 0.47 and 0.95, respectively. 

Figure \ref{F1-QR} displays the periodic quantum oscillations of $G$ (\ref{G-spinres}), \textit{versus} the  AC phase $\phi_{\textsc {so}}$ in the interval $[0,2]$ for different values of transmission coefficients at the lead-to-ring junctions $T_{o}(\varepsilon)$, together with QPC$_3$ fully open. The latter implies $T_{s}(\varepsilon)=1$, which entails $\hbar\omega_{j} = 0.2$ meV   and $\phi = 0.5$ for the AB-phase.  Moreover the dynamic phase  has been set to $\theta=\pi k_{\textsc f}R  = 39$, with  $k_{\textsc f} = \sqrt{2\pi n_{e}}$ and $n_{e} = 3.93 \times 10^{10}$ cm$^{-2}$. The panel of Fig.4.{\it a} correspond to a situation in which both QPC$_{1,2}$  are fully open ($T_{o}=1$), so there is no scattering of the moving electrons around the QR. The oscillations of the conductance with $\phi_{\textsc {so}}$ are a clear manifestation of the  spin-interference. As described in {\it Case (i)} above, the maximums (resonances) of the conductance are present always that  $\phi_{\textsc {so}}^{max} = n$, here with $n=0,1,2$ whereas the minimum (anti-resonances) appear as long  $\phi_{\textsc {so}}^{min} = \frac{n}{2}$, with $n=1,3$. As soon QPC$_{1,2}$ are set to partial transparency ($T_{o}\ne 1$), the pattern of the quantum oscillations of the conductance changes drastically and the resonances  follows Eq.(34) producing double peaks around each anti-resonance which now appears at $\phi_{\textsc {so}}^{min} = \frac{n}{2}$, with $n$-integer.

To further analyze the results let us consider the particular case of Fig.4c in which both QPC$_1$ and QPC$_2$ are set to half transparency ($T_o(\varepsilon)=R_o(\varepsilon)=1/2)$ while QPC$_3$ is totally open ($T_s=1)$. From Eq.(34) the resonances of the conductance  are governed by
\begin{equation}
 \label{G-max4}
  \phi_{\textsc {so},\pm}^{max} =n \pm\frac{1}{2\pi}\left[c_1+c_2 {\text{ arcCos}}(\cos2\theta) \right],
\end{equation}
\noindent with $n$ integer, $c_1=17-12\sqrt{2}$ and $c_2=1-c_1$. On the other hand the the minimums of the conductance arises here as long $\phi_{\textsc {so}}^{min} = \frac{n}{2}$, with $n$-integer (see Table I). Clearly as we would expect, the width of the resonances diminishes as the opacity of the QPC$_{1,2}$  increases. In the limit of vanishing transparency, when $T_{o}(\varepsilon) = 0.001$ (see Fig.\ref{F1-QR}f), each QPC at the lead-to-ring junctions have thus the largest opacity. In this scenario a low-probability flux through the lead-to-ring junctions is allowed -despite the strong confinement potential imposed- given rise to the sharp resonances observed as the eigenenergies of the QR are being mapped.

\begin{figure}
 \hspace*{0.0in}\includegraphics[width=6.0in,height=5.5in]{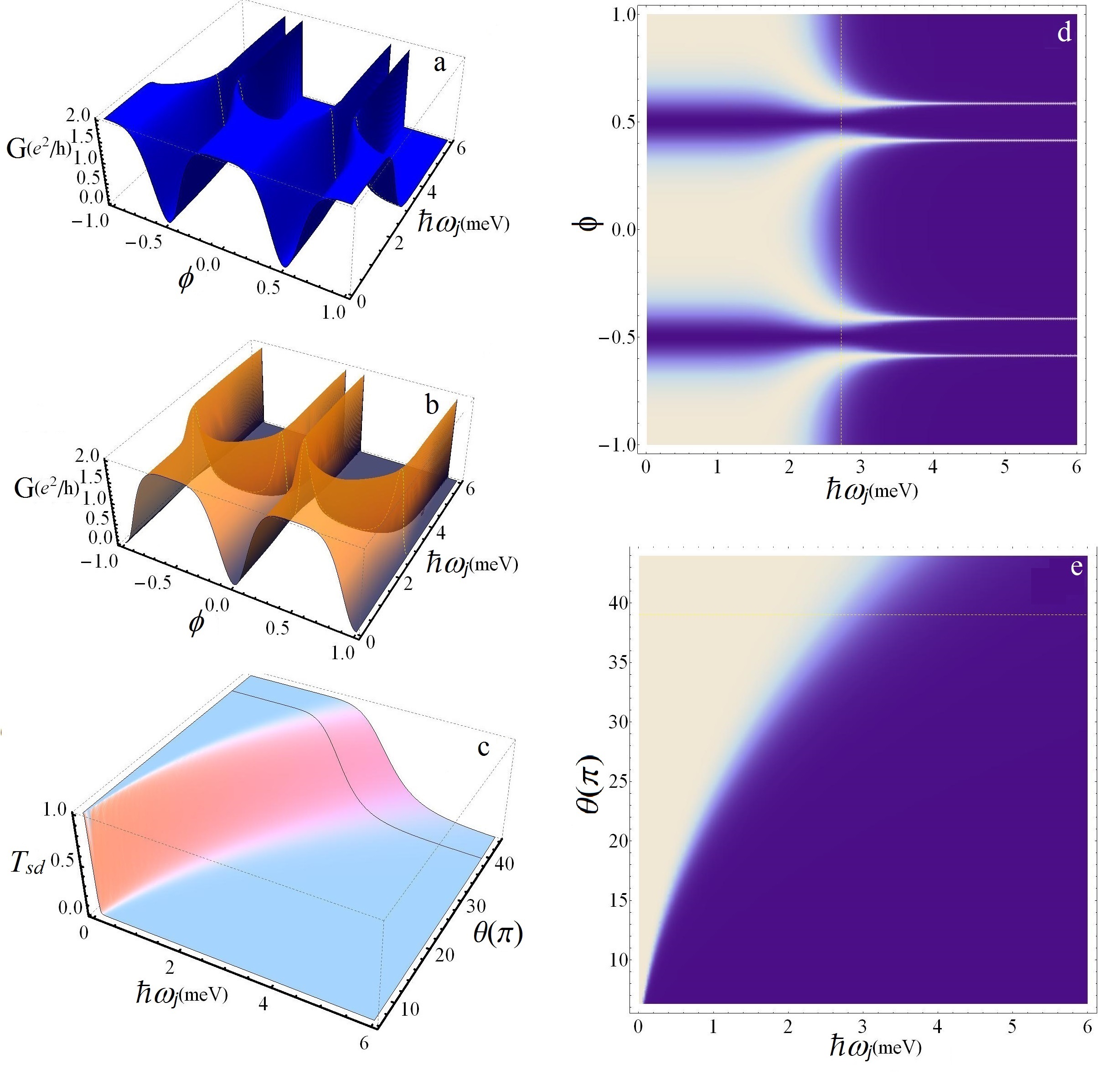}
\vspace{-0.15in}
  \caption{\label{F3-QR} (Color online) Panels (a),(b) and (d) show the oscillations of $G$ as a function of the AB phase $\phi$ and confinement energy $\hbar\omega_{j}$. For plot in panels (a) and (d), we use $\phi_{so} = 0.5$ ($\alpha_{so}=21$ meV\,nm), while in panel (b) we have set $\phi_{so} = 1.0$ ($\alpha_{so}=42$ meV\,nm). The dotted vertical line in panel (d), points the value $\hbar\omega_{j} = 2.72$ meV (equivalent to $T_{j} = 1/2$). The panels (c) and (e) present the oscillations of  transmission coefficient (\ref{T-Fertig}), as a function of $\hbar\omega_{j}$ at the junctions, and the phase $\theta$. The limit of the Fermi dynamic phase $\theta_{\textsc f} \approx 39.2$ (solid black/dashed white line) is indicated in panel (c)/(e). The light/dark distribution of the density maps corresponds to the higher/smaller values of  $G$, respectively.}
\end{figure}

In Figure \ref{F3-QR}.a\textcolor[rgb]{0.00,0.00,1.00}{(b)} we show plots of the quantum conductance of the QR with QPC$_3$ fully transparent ($T_{s} = 1$) and the other QPC$_{1,2}$ partially open, as a function of AB-phase $\phi$ and confinement strength $\hbar\omega_{j}$, in the range between [-1,1] and [0,6] meV, respectively. The plot in panels (a) and (b) show $G$ for two values of AC-phase ($\phi_{\textsc {so}} = 0.5,1.0$, respectively). Notice the shifting of the $G$-oscillation pattern as a function of $\phi$ by tuning selectively two values for the SOI parameter $\alpha_{\textsc {so}}$. The behavior of the conductance with $\hbar\omega_{j}$ exhibits a clear crossover around  $\hbar\omega_{j} = 2.7$ meV regardless of the AB-phase value. Indeed, for $\hbar\omega_{j} < 2.72$ meV (equivalent to $T_{j} = 0.5$) the conductance {\it versus} $\phi$ shows the expected widespread periodic AB-oscillations\textcolor[rgb]{0.00,0.00,1.00}{\cite{Buttiker84,ShelykhPRB72}} with maximum amplitude $2e^2/h$. However for confinement strengths $\hbar\omega_{j} > 2.72$ meV  the quantum transport through the QR device is strongly influenced by the opacity of QPC$_{1,2}$ (larger coupling lead-to-ring) and rapidly reduces to vanishing values. As a consequence the widespread periodic oscillations of $G$ evolves into very sharp resonant features. This behavior can be seen more clearly in panel $5.\textcolor[rgb]{0.00,0.00,1.00}{d}$ showing  a density map of $G$ against the plane ($\phi,\hbar\omega_{j}$). The white dotted vertical line at the transition regime $\hbar\omega_{j} = 2.72$ meV is drawn here just to guide the eye.

It is illustrative to have a close look to the model transmission coefficient (\ref{T-Fertig}) for the QPCs, as a function of $\hbar\omega_{j}$ and  the dynamic phase $\theta$.  This is the aim of the Fig.\ref{F3-QR}c \textcolor[rgb]{0.00,0.00,1.00}{(e)}. The transmission probability $T_{\textsc {sd}}$, is adiabatically modulated following a smooth step-like profile from its maximum value ($T_{\textsc {sd}}=1)$ at its plateau, to essentially zero at relatively large values of $\hbar\omega_{j}$ and $\theta$. As a result the coherent quantum interference phenomena and the electronic flux in the QR device is affected accordingly as $T_{\textsc {sd}}$ in QPC$_{1,2}$ is modulated. For instance,  the transmission probability at the neighborhood of $\theta_{\textsc f} \approx 40$ ({\it i.e.} $E_{\textsc f} \approx 43.2$ meV), indicated with a solid black (white dashed) line in Fig.\ref{F3-QR}c\textcolor[rgb]{0.00,0.00,1.00}{(e)}, changes significantly as $\hbar\omega_{j}\,$ is varied in the interval between $2.4$  and $3.6$ meV. Outside this range, the values of $T_{\textsc {sd}}$ are basically uniform; being maximal for $\hbar\omega_{j} < 2.4$ meV ($T_{\textsc {sd}} \rightarrow 1$), and minimal ($T_{\textsc {sd}} \rightarrow 0$) when  $\hbar\omega_{j} > 3.6$ meV (see Fig.\ref{F3-QR}\textcolor[rgb]{0.00,0.00,1.00}{e}). Such behavior of $T_{\textsc {sd}}$   at each lead-to-ring junction (QPC$_{1,2}$) explains well the two drastic different features of the $G$-oscillations with respect to $\hbar\omega_{j}$; from the widespread of the periodic $G$-oscillation, to the resonant response, as shown in Fig.\ref{F3-QR}{\it a}({\it b})\textcolor[rgb]{0.00,0.00,1.00}{({\it c})}.

We now turn to another interesting phenomenology. This concerns  the situation in which the $G$-oscillation pattern (as a function of $\phi$) are shifted uniformly just by tuning selectively the Rashba SOI  parameter $\alpha_{\textsc {so}}$. These are represented in Fig.5{\it a} as the two superimpose  conductance plots corresponding to the two different values for the AC-phase, $\phi_{\textsc{so}}$ = 0.5(1.0).
From Eqs. (31) and (32) it is easy to show that maximums in $G$ should appear as long $\phi^{max}_{\pm} = \frac{1}{2}\mp  \phi_{\textsc {so}}-n$, with $n$ integer.
Therefore the dashed horizontal lines in panel ({\it b}) at $\phi = \pm 0.5$ values, would correspond to the conductance maximums [dark curve in panel ({\it a})] for which $\phi_{\textsc {so}} = 1.0$, while $G$-oscillations of the $\phi_{\textsc {so}} = 0.5$ [lighter  curve in panel ({\it a})] case appears shifted by a phase of $0.5$ as we would expect.

\begin{figure}
 \hspace*{0.0in}
 \includegraphics[width=6.0in,height=5.8in]{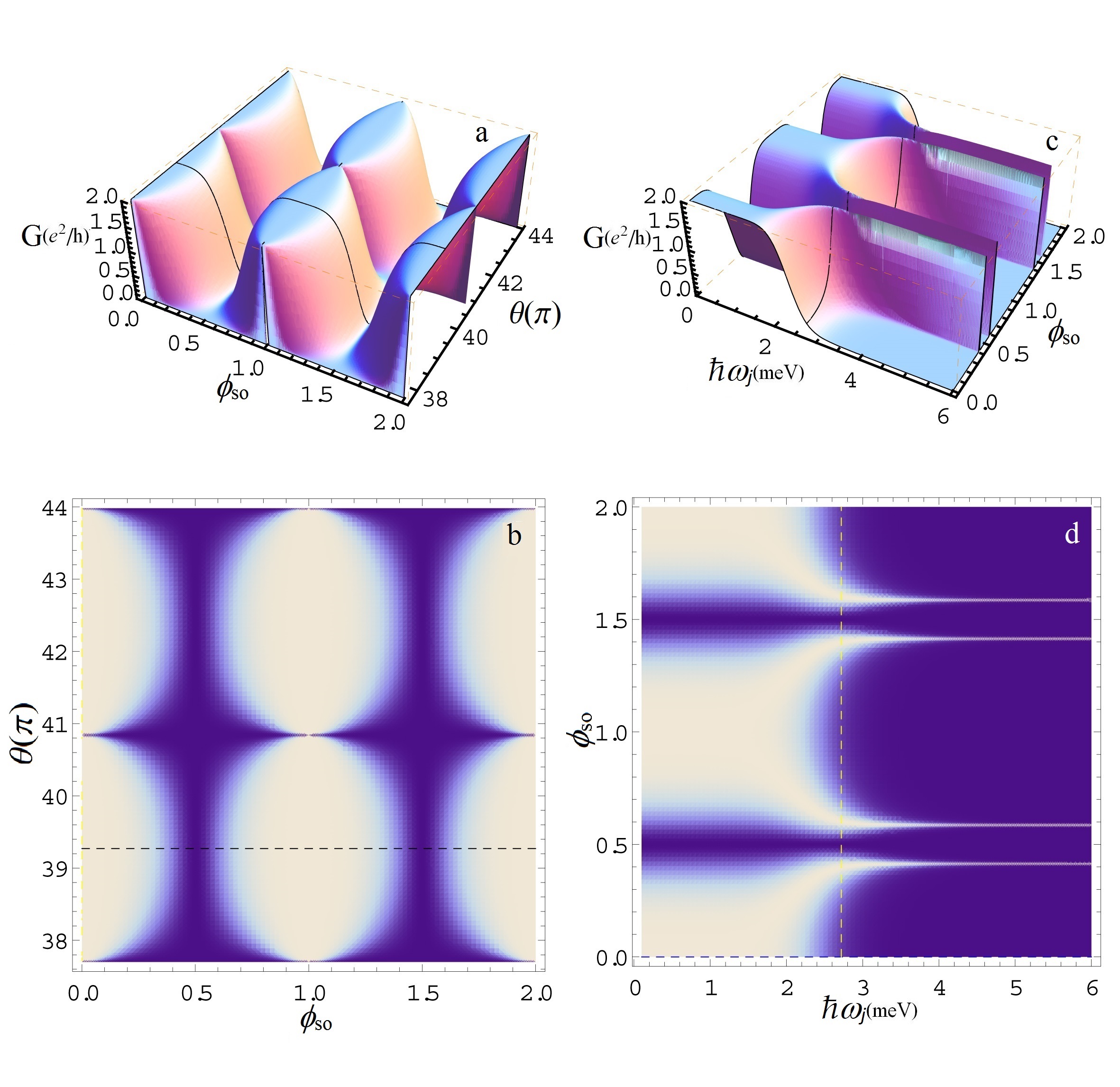}
\vspace{-0.3in}
 \caption{\label{F2-QR} (Color online) Panels {\it a}) and ({\it b}) shows $G$-oscillations, as a function of the AC phase $\phi_{\textsc {so}}$ and the dynamic phase $\theta$. The limit of the Fermi dynamic phase $\theta_{\textsc f} \approx 39.2$ (dashed dark line), is indicated.  The panel ({\it c})/({\it d}) presents the same, as a function of the confinement $\hbar\omega_{j}$ at the junctions, and the phase $\phi_{\textsc {so}}$ in a density map for $G$. The limit of $\hbar\omega_{j} = 2.72$ meV (dashed light line) is indicated. The light-dark distribution within the density map corresponds to a strong-weak gradient for the values of  $G$. All plots were calculated at a magnetic field $B=10$ mT, corresponding to a AB-phase of $\phi=0.5$ for the studied QR. }
\end{figure}

In Figures \ref{F2-QR}.{\it a}({\it b}) we present the plots of the  evolution of the quantum conductance oscillations  of the QR-device as a function of the AC-phase $\phi_{\textsc {so}}$ and the dynamic phase $\theta$, for a fix AB-phase of $\phi=0.5$ and confinement strength of $\hbar\omega_{j}=0.2$ meV (which imposes $T_{o} =1$ for QPC$_{1,2}$), whiles setting QPC$_3$ at its maximum transparency ($T_{s} =1$). The periodic oscillatory patterns found for $G$ against $\phi_{\textsc {so}}$  respond the to AC spin-interference effects induced by the SOI-R, as had been earlier predicted and measured for biased AC-QRs at low temperature.\textcolor[rgb]{0.00,0.00,1.00}{\cite{Nitta99,Vasilopoulos04,Shelykh05-2, Nitta03}} On the other hand, the periodicity of $G$ against the dynamic phase $\theta$ comes  from the ${\sin}^2\theta$  dependence of the spin-resolved conductance in Eq.(\ref{G-spinres}) and arises due the coherent quantum interference of the winding electrons around the QR.

The oscillatory pattern changes drastically when a variable lateral confinement strength of QPC$_{1,2}$ is considered while maintaining QPC$_3$ fully open. This is shown in Fig.\ref{F2-QR}.{\it c}({\it d}) for a fix dynamic phase of  $\theta = 39$. Similar as it occurs in Fig.\ref{F3-QR}{\it c}({\it d})  the evolution of the quantum oscillations of $G$ against the AC-phase reaches a crossover  into a well-defined abrupt resonances as the QPC confinement energy $\hbar\omega_{j}$ is increased. The overall phenomenology shown here is identical of what arises in   Fig.\ref{F3-QR}{\it c}({\it d}) by just replacing $\phi \leftrightarrows \pm\phi_{\textsc {so}}$.  This is expected by the symmetry of the AC and AB phases as can be verified from  Eq.(32).

\begin{table}
\caption{\label{TabExtrem} Extremes of (\ref{G-max2}) obtained from Eq.(36) for a few values of $n$.
We have fixed $T_{o} = 0.5$, and $\theta = 39.$}
\begin{ruledtabular}
\begin{tabular}{rccc}
$n$ & $ \,\,\,\,\,\,\phi_{\textsc {so}\pm}^{min}$ & $ \phi_{\textsc {so-}}^{max }$ & $ \phi_{\textsc {so+}} ^{max }$  \\
\hline \hline
$-2$ & $-1$ & $-2.39841$ & $-1.60159$\\
$-1$ & $-0.5$ & $-1.39841$ & $-0.60158$ \\
$\,0$ & $\,\,\,0$ & $-0.39841$ &\,\, $0.39841$ \\
$1$ & $\,\,\,0.5$ & $\,\,\,\,  0.60158 $ &\,\, $1.39841$ \\
$2$ & \,\,\,1 & $\,\,\,\,1.60159$ &\,\, $2.39841$ \\
\end{tabular}
\end{ruledtabular}
\end{table}

\begin{figure}
 \hspace*{-5mm}
 \includegraphics[width=6.0in,height=5.8in]{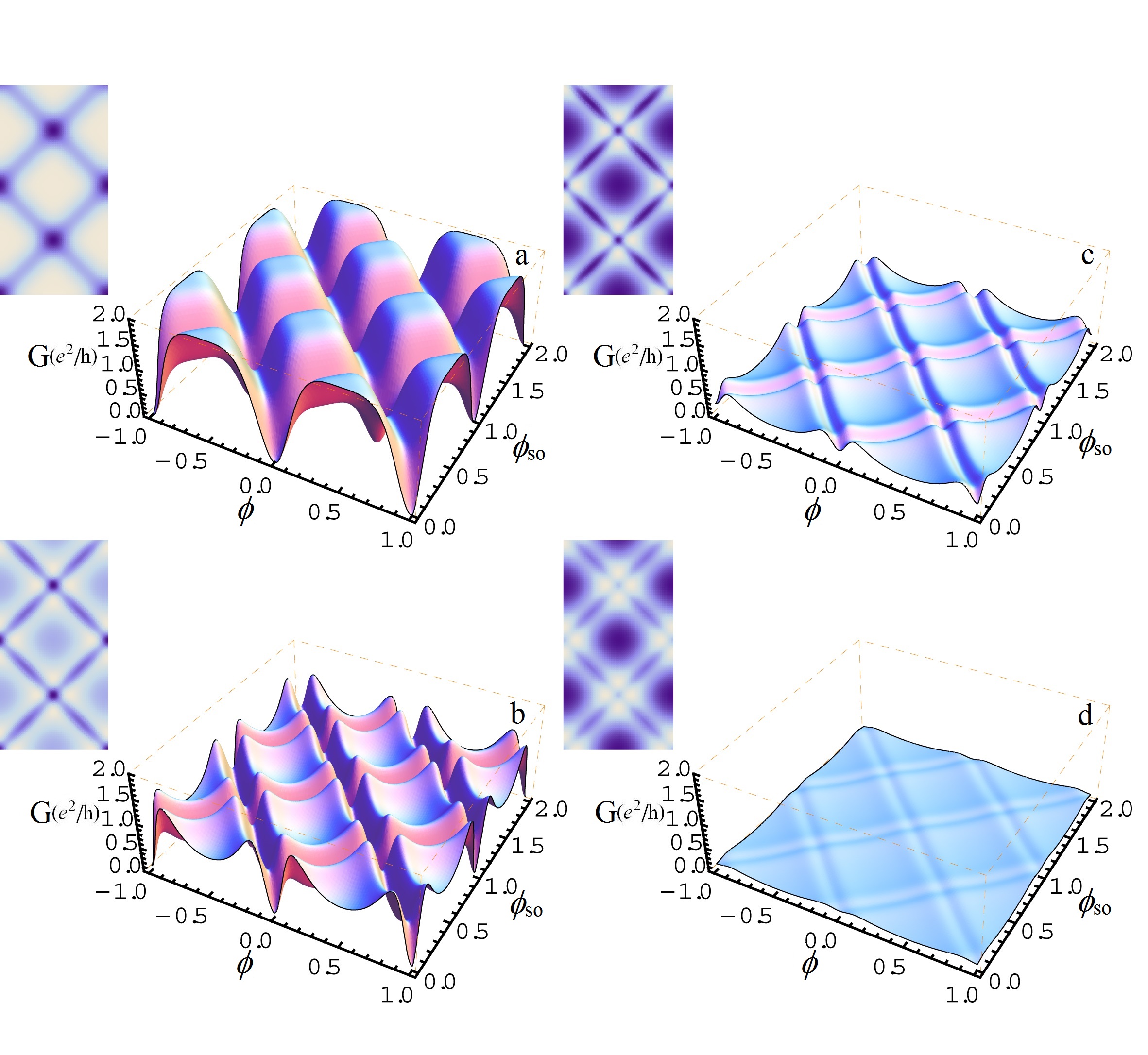}
\vspace{-0.3in}
 \caption{\label{F5-QR} (Color online ) Conductance oscillations as a function of the AB phase $\phi$ and AC phase $\phi_{\textsc {so}}$. The transmission coefficients used at the junctions (QPC$_{1,2}$) and at the scatter (QPC$_3$), in the panels {\it a, b, c, }and {\it d}; correspond to: $T_{o} \equiv T_{s} = 1, 3/4, 1/2$, and $1/3$, respectively. Since $T_{o(s)}$ depend also on $\hbar \omega_j$, these are varied simultaneously as the AB-phase $\phi$ is tuned  in order to have an uniform transmission probability of the QPCs as $\phi$ changes.  The insets show the same information in a density-map format for $\phi$ ranging in the interval $[-1,0]$ which corresponds to a magnetic field interval between -21 to 0 mT.}
\end{figure}

The influence of the transmitivity of   QPC$_3$ in the conductance of the QR device assuming identical transmitivity to that of the other two QPCs at  the lead-to-ring junctions is investigated in Fig.\ref{F5-QR}.  The conductance $G$ is calculated as a function of phases $\phi$ and $\phi_{\textsc {so}}$ for four different  values of the transmission probability of the QPC$_{1,2,3}$ ($T_{o} = T_{s}=1, 3/4, 1/2$, and $1/3$) and plotted in Fig.\ref{F5-QR}.{\it a-c}, respectively. In Fig.\ref{F5-QR}{\it a} the well known oscillating periodic response of   $G$ for a QR without scatterers ({\it i.e.} in the absence of QPCs) are nicely reproduced here. For instance, similar behaviors had been reported for an AB ring without any scatter in the arms of the QR.\textcolor[rgb]{0.00,0.00,1.00}{\cite{Vasilopoulos04}}  The periodic plateau formation at $2e^{2}/h$  with the AB and AC phases  in Fig.\ref{F5-QR}{\it a}  are also in agreement with earlier results for   InGaAs/InAlAs QR structures.\textcolor[rgb]{0.00,0.00,1.00}{\cite{ShelykhPRB72}} The new behavior in the QR device arises here as soon a finite scattering of the traveling electrons is considered due the presence of partially open QPCs at the lead-to-ring junctions and in the upper arm of the QR. As a consequence of effectively closing the QPC$_{1,2,3}$ the observed plateaus of $G$ (in units $2e^{2}/h$) shown in Fig.\ref{F5-QR}{\it a}  at multiple half-integer values of $\phi$ and $\phi_{\textsc {so}}$, evolves adiabatically to depression-like structures, forming periodic crater-type features instead, see Fig. \ref{F5-QR}{\it b-c}. Due the reduced flux of electrons allowed to enter the QR by diminishing  $T_{o}$ and $T_{s}$, a rather sensitive and strong tendency to fade away the amplitude of the overall  conductance response is predicted, as being shown in panel $d$ Fig.\ref{F5-QR}.  Notice that, the robustness of the periodic pattern of $G$  are still visible even though the transmission of the QPCs are  strongly reduced up to $1/3$ [see Fig.\ref{F5-QR}({\it b-d}) and insets].

\begin{figure}
 \includegraphics[width=6.0in,height=4.3in]{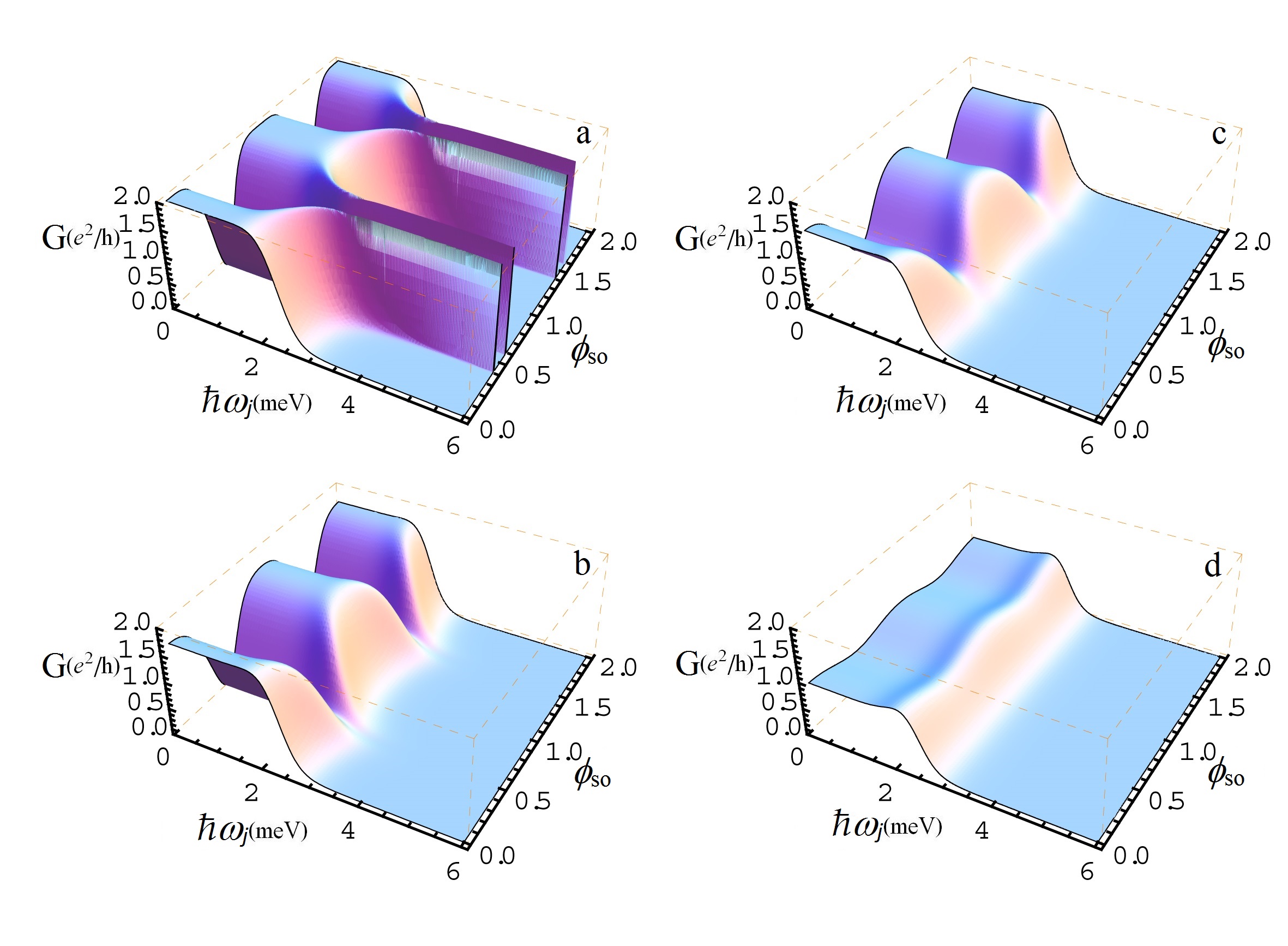}
\vspace{-0.3in}
  \caption{\label{F6-QR} (Color online) Behavior of the conductance as a function of $\hbar\omega_{j}$ and AC phase $\phi_{\textsc {so}}$. We have fixed the AB phase $\phi = 0.5$ ($B=10$ mT). The transmission coefficients used at the scatter (QPC$_3$), in the panels a, b, c, and d correspond to $T_{s} = 1, 1/2, 1/4$ and $10^{-3}$, respectively. }
\end{figure}

Thus, just by controlling simultaneously the aperture of QPC$_{1,2,3}$  we can achieve three qualitative different physical configurations for conductance response of the QR device as described in panels $a$, $b$ and $c$ of Fig.\ref{F5-QR}. Notice also that due the possible control of the Aharanov-Casher phase by the electrical tuning of the Rashba spin-orbit coupling together with the gate control of the confinement of the QPCs, leads to the possibility of switching ON and OFF the current through the QR device, even in the absence of magnetic field ($\phi =0$). A similar scenario can be achieved by, on the contrary, fixing the Rashba coupling and varying the QPCs confinement strength, also at zero magnetic field. This constitutes an  advantageous feature over those QR devices that rely on a magnetic flux to manifest a similar behavior.
This quantum effect suggests the possibility of using our QR device proposal as a logic gate\textcolor[rgb]{0.00,0.00,1.00}{\cite{Foldi05}} and in appealing spintronics applications devices whose performance can be manipulated by adjusting gate voltages only.

Finally, in Fig.\,\ref{F6-QR} we study the behavior of the conductance quantum oscillations  as a function of $\hbar\omega_{j}$ and AC phase $\phi_{\textsc {so}}$ for several values of the transmission coefficients $T_s$ for the upper-arm  QPC$_3$. Cases with $T_s=1,1/2,1/4$ and $10^{-3}$ are considered in panels ${\it a,b,c}$ and ${\it d}$, respectively.
It is clear that the persistent resonances appearing for $\hbar\omega_{j} \gtrsim 3 $ meV in Fig.8{\it a} vanish completely by closing QPC$_{3}$, see Fig.8{\it b} and $c$ for instance. Such effect respond to the disappereance of the spectrum of the quasi-stationary states of the close QR as soon a sizable scatterer is present in the path way of the circling electrons, which in turn destroys quantum interference. Notice that despite   the maximum opacity ($T_s=10^{-3}$) imposed in the QPC$_{3}$ (Fig.\ref{F6-QR}{\it d}) the conductance do not vanish completely but rather  fluctuates around $G \approx e^{2}/h$, in  sharp contrast with the phenomenology observed in Fig.\ref{F5-QR}{\it a/d }in which the QPCs at the lead-to-ring-junctions are essentially blocked instead. In the former case the coherent quantum interference of electrons is completely suppressed.

These features described above comprises yet another distinctive behavior of $G$ profiles in the QR device proposed here. Indeed,  depending on the value of $T_{s}$, and  by considering the full axis $ 0\leq\hbar\omega_{j}\leq6$ meV, one can obviously achieve three different configurations for $G$. Namely, as shown in Fig.\ref{F6-QR}{\it a} maximum fluctuations of $G$ can be attained in conjunction with very sharp resonances, while  $e.g.$ in panel ({\it c}) the wide oscillations are diminished in amplitude and the sharp resonance vanishes. The third distinctive pattern is seen in Fig.\ref{F6-QR}{\it d}, where the oscillations are strongly suppressed due the strong confinement strength in QPC$_3$ which in turn inhibit spin quantum interference.

\section{Summary}
\label{Sec:Summary}

We have examined the electron spin-interference and the coherent electronic spin-transport through a proposed semiconductor QR-device with three embedded QPCs in the presence of Rashba spin-orbit interaction. A suitable model for the QPCs based in two-dimensional saddle point potentials and known transmission coefficients is employed. In contrast to previous theoretical studies, the quantum point contacts are included at the inlet and outlet of the QR, which can be used to control the transmission in different parts of the structure and give us the freedom to vary electrically the interference pattern of the QR. 
Using the $\mathbb{S}$-matrix formalism we have derived a closed analytical expression for the total transmitted amplitude through the QR-device  that incorporates the confinement strength, external magnetic field and Rashba spin-orbit coupling in the same footing. We also derive the conditions for resonances and anti-resonances of the conductance of the QR. Such formulas for the spin-conductance holds in general, independent of the transmission probability model considered for the QPCs. 
Our theoretical modeling for the Landauer conductance  readily reproduce the expected  periodic-flux conductance effects reported in the literature in the absence of QPCs; with and without Rashba coupling. 

When considering the influence of the scattering at the QPCs, we found that the electron spin  interference in the QR-device is very sensitive to the QPC confinement strengths, offering new features amenable to control the electron spin-transport and the quantum spin-interference in the QR. For instance we found that, ({\it i}) the oscillatory pattern and the spin interference in the QR can be controlled to great extend by varying the lead-to-ring transmission probabilities (through the confinement strengths of the QPCs), ({\it ii}) the conductance oscillations exhibit a crossover to well-defined sharp resonance behavior as the QPCs confinement is increased, ({\it iii}) the simultaneous electrical control of the aperture of all QPCs yields three remarkably different behavior of the conductance response as a function of the Aharanov-Bohm and Aharanov-Casher phases,  and ({\it iv}) also distinctive behavior of the conductance results by varying the aperture of the upper-arm QPC and the Rashba spin-orbit coupling. 

The new features predicted in the proposed QR-device may be of utility to implement spintronic logic gate devices based on quantum interference and whose function can be manipulated using gate voltages only.  The theoretical modeling is flexible to incorporate more scatterers in the QR, Dresselhaus type of spin-orbit interaction\textcolor[rgb]{0.00,0.00,1.00}{\cite{Shakouri}}, spin-flip mixing and asymmetric injection/detection of electrons.

\section*{Acknowledgments}
 One of the authors (L.D-C) is grateful to the Visiting Academic Program of the UIA-M\'{e}xico and acknowledges the hospitality and  facilities offer by CNyN-UNAM, Ensenada, M\'{e}xico where part of this research was executed.
This work was partially supported by project PAPIIT-DGAPA (UNAM) No. IN109911.

\appendix
\section{Derivation of the total transmission amplitude}

In this appendix we derive Eq.(28) for the scattering amplitude $\alpha\p_{2\sigma}$ that characterize the probability of the emergent electrons to the right hand side of the QR device.

First we realize that with the aid of the expressions (\ref{tra-ref(31)}),(\ref{tra-ref(32)}) and (\ref{tra-ref(3)}), it is possible to establish the useful  relationship
\begin{eqnarray}
\label{tra-ref(1Up2)}
   \left[
    \begin{array}{l}
      \beta_{2\sigma} \\
      \beta\p_{2\sigma}
    \end{array}
  \right] =  \mathbf{\Phi}^{\sigma}_{u} \cdot \mathbb{T}_{u} \cdot \mathbf{\Phi}^{\sigma}_{u}
  \left[
    \begin{array}{l}
      \beta\p_{1\sigma} \\
      \beta_{1\sigma}
    \end{array}
  \right],
 \label{tra-ref7-4}
\end{eqnarray}

\noindent then consider (\ref{tra-ref(Lw)}) in the compact form
\begin{equation}
   \left[
    \begin{array}{l}
      \gamma_{1\sigma} \\
      \gamma\p_{1\sigma}
    \end{array}
  \right]  =
    \mathbf{\Phi}^{\sigma}_{l} \cdot  \mathbb{T}_{l}
   \left[
    \begin{array}{l}
      \gamma\p_{2\sigma} \\
      \gamma_{2\sigma}
    \end{array}
  \right],
\end{equation}
\vspace{-14mm}
\begin{flushright}

\end{flushright}
\noindent and  assume electron incidence from the left lead only, with an unitary incident current flux, {\it i.e.} $\alpha_{1\sigma} =1$ and $\alpha_{2\sigma}=0$. By using (A1) and (A2) together with (20)-(23), (25) and (28) it is possible to connect the amplitudes of the ingoing/outgoing waves arriving/leaving the QPC$_{1}$ and QPC$_{2}$ in  the QR by writing
\begin{equation}
   \left[
    \begin{array}{l}
      \beta\p_{1\sigma} \\
      \beta_{1\sigma}
    \end{array}
  \right]  =
  -\frac{\sqrt{2}\,T_{s}^{1/2}}{\left(R_{s}^{1/2}+\lambda \right)}
   \left[
       \begin{array}{c}
         \lambda \\
         -1
       \end{array}
    \right] + \bn{t}_{j}
   \left[
    \begin{array}{l}
      \gamma_{1\sigma} \\
      \gamma\p_{1\sigma}
    \end{array}
  \right],
\end{equation}
\vspace{-14mm}
\begin{flushright}

\end{flushright}
\noindent whereas
\begin{equation}
   \left[
    \begin{array}{l}
      \gamma\p_{2\sigma} \\
      \gamma_{2\sigma}
    \end{array}
  \right]  =   \bn{t}_{j}
    \left[
    \begin{array}{l}
      \beta_{2\sigma} \\
      \beta\p_{2\sigma}
    \end{array}
  \right],
\end{equation}
\vspace{-14mm}
\begin{flushright}

\end{flushright}
\noindent being $\lambda = \textsf{c}-\textsf{b}= \pm 1 $. Next, by following a similar procedure of Ref.\textcolor[rgb]{0.00,0.00,1.00}{\onlinecite{Vasilopoulos07}} we have derived
\begin{equation}
  \bn{t}_{j1} \equiv \bn{t}_{j2} = \bn{t}_{j} =  \frac{1}{\textsf{c}}
   \left(
   \ba{cc}
     \textsf{c}^{2} - \textsf{b}^{2} & \textsf{b} \\
       -\textsf{b} & 1
   \ea
  \right),
\end{equation}
\vspace{-14mm}
\begin{flushright}

\end{flushright}
\noindent  where $\bn{t}_{j1}$, and  $\;\bn{t}_{j2}$  represents the effective transmission amplitudes matrices at the left and right lead-to-ring junctions, respectively. From Eqs. (A1), (A2) and (A3) we obtain
\begin{equation}
  \left[
     \begin{array}{l}
      \beta\p_{1\sigma} \\
      \beta_{1\sigma}
    \end{array}
  \right]  =
  -\frac{\sqrt{2}\,T_{s}^{1/2}}{\left(R_{s}^{1/2}+\lambda\right)}
   \left[
       \begin{array}{c}
         \lambda \\
         -1
       \end{array}
    \right] + \bn{t}_{j}\mathbf{\Phi}^{\sigma}_{l}\mathbb{T}_{l}\bn{t}_{j}
   \left[
    \begin{array}{l}
      \beta_{2\sigma} \\
      \beta\p_{2\sigma}
    \end{array}
  \right].
\end{equation}

Similarly, we can conveniently write the amplitude of the outgoing wave to the right lead as follows
\begin{eqnarray}
   \left[
    \begin{array}{c}
     \alpha\p_{2\sigma} \\
     0
    \end{array}\right] = -\left(\frac{\sqrt{2}T_{s}^{1/2}}{ R_{s}^{1/2}+\lambda } \right )
 {\bn \lambda}\left[
    \begin{array}{l}
       \beta_{2\sigma} \\
      \beta\p_{2\sigma}
    \end{array}
  \right],
\end{eqnarray}
\vspace{-14.5mm}
\begin{flushright}

\end{flushright}
\noindent where we have used (29)
\begin{equation}
 {\bn \lambda}=\left(
    \begin{array}{cc}
      \lambda & 1\\
      0 & 0
       \end{array}
  \right),
\end{equation}
\vspace{-15.5mm}
\begin{flushright}

\end{flushright}

\noindent in which either value of  $\lambda (\pm 1)$ gives a possible solution of the scattering problem.  We can now use  (A6) and (A7) we arrive to the column vector
\begin{equation}
  \left[
    \begin{array}{l}
      \beta_{2\sigma} \\
      \beta\p_{2\sigma}
    \end{array}
  \right] =  \left(\frac{\sqrt{2}\,T_{s}^{1/2}}{R_{s}^{1/2}+\lambda}\right)
  (\mathbb{P}^{\,\sigma})^{-1}\mathbf{\Phi}^{\sigma}_{u}\mathbb{T}_{u}
   \mathbf{\Phi}^{\sigma}_{u}
     \left[
       \begin{array}{c}
         \lambda \\
         -1
       \end{array}
    \right],
\end{equation}
\vspace{-12mm}
\begin{flushright}

\end{flushright}
\noindent in which we have introduced the ($2\times 2$) matrix $\mathbb{P}^{\,\sigma} = \mathbf{\Phi}^{\sigma}_{u} \mathbb{T}_{u}\mathbf{\Phi}^{\sigma}_{u}\bn{t}_{j1}\mathbf{\Phi}_{l}\mathbb{T}_{l}\bn{t}_{j2} - \bn{I}_{2}$.  Finally, inserting (A9) into (A7) we get (28)
\begin{equation}
   \left[
    \begin{array}{c}
      \alpha\p_{2\sigma} \\
      0
    \end{array}
  \right]  =
 \frac{2 T_{s}}{\left( R_{s}^{1/2}+\lambda \right)^2}
    {\bn \lambda}\,(\mathbb{P}^{\,\sigma})^{-1}\mathbf{\Phi}^{\sigma}_{u}\mathbb{T}_{u}
   \mathbf{\Phi}^{\sigma}_{u}
    \left[
       \begin{array}{c}
         -\lambda \\
         1
       \end{array}
    \right],
\end{equation}
\noindent which provides a closed form for the total transmitted amplitude $ \alpha\p_{2\sigma}$ of the outgoing electrons at the right lead in terms of the transmission amplitudes of the QPCs and all the quantum phases involved.

\end{document}